\documentclass{aa}  

\usepackage{graphicx}
\usepackage{txfonts}
\usepackage{lipsum}
\usepackage{subcaption}         
\usepackage{lscape}             
\usepackage{placeins}          
\usepackage[font=small]{caption}
\usepackage{bm}
\usepackage{array}
\usepackage{tablefootnote}
\usepackage{tabularx}
\usepackage{booktabs}
\usepackage{threeparttable}
\usepackage{textcomp}              
\usepackage[colorlinks=true,     linkcolor=blue, citecolor=blue, filecolor=blue, urlcolor=blue]{hyperref}
\usepackage[normalem]{ulem}
\newcommand{\dpa}{\Delta_{\mathrm{PA}}}
\newcommand{\tage}{\tau^{\mathrm{\change{proxy}}}}

\newcommand{\taum}{\tau^{\mathrm{meas}}}
\newcommand{\linj}{L^{\mathrm{\change{proxy}}}}
\newcommand{\dnion}{\frac{\mathrm{d}n_{\mathrm{He^{+}}}}{\mathrm{d}t}}
\newcommand{\dnionp}{\frac{\mathrm{d}n_{\mathrm{He^{+}}}}{\mathrm{d}t^\prime}}
\newcommand{\cts}{C_{\mathrm{He}^{+}}}
\newcommand{\ctstorus}{C^{\mathrm{torus}}_{\mathrm{He}^{+}}}
\newcommand{\nn}{n_{\mathrm{He}}}
\newcommand{\Ir}[1]{I_{\mathrm{He}, #1}}
\newcommand{\I}{I_{\mathrm{He}}}
\newcommand{\vc}{v_{\mathrm{conv}}}
\newcommand{\vsw}{v_{\mathrm{sw}}}
\newcommand{\niontot}{n_{\mathrm{He}^+}^{\mathrm{total}}}
\newcommand{\nionprod}{n_{\mathrm{He}^+}^{\mathrm{prod}}}
\newcommand{\niontorus}{n_{\mathrm{He}^+}^{\mathrm{torus}}}
\newcommand{\nninf}{n_{\mathrm{He}}^\infty}

\newcommand{\nionpart}{n_{\mathrm{He}^+}^{\mathrm{partial}}}

\usepackage{tikz}
\newcommand\wye[1][]{%
    \tikz\draw[thick, line cap=round,x=1ex,y=1ex,#1]
    (0,0) -- ++(90:1)
    (0,0) -- ++(-30:1)
    (0,0) -- ++(-150:1);
}
\newcommand{\change}[1]{#1}
\begin{document}
   \title{
                Observations of stable pickup $\mathrm{He}^{+}$ tori in a magnetic flux rope at 0.85 au}

        \author{
                Chaoran Gu
                \inst{1}\corrauth{chaorangu@physik.uni-kiel.de}
                \and
                Lars Berger\corrauth{berger@physik.uni-kiel.de}
                \inst{1}
                \and
                Verena Heidrich-Meisner
                \inst{1}
                \and
                Erik Jentsch
                \inst{1}
                \and
                Robert F. Wimmer-Schweingruber
                \inst{1}
                \and
                Lars Seimetz
                \inst{1}
        }

        \institute{Institut für Experimentelle und Angewandte Physik (IEAP), Christian-Albrechts-Universität zu Kiel, Leibnizstr. 11, 24118 Kiel, Germany
        }
        \date{unknown}

  \abstract
   {Interstellar pickup ions originate from the neutral interstellar medium, enter the heliosphere, are ionized and picked up by the solar wind. They initially form an anisotropic torus-shaped velocity distribution function, which is generally believed to be transformed into an isotropic shell distribution by pitch-angle scattering. The time scale of pitch-angle scattering is still under debate and is often assumed to be instantaneous. Limited by instrumental capabilities and potentially fast pitch-angle scattering, torus-shaped velocity distribution functions are not commonly observed. We present observations of clear torus-shaped velocity distribution functions in a magnetic flux rope at an unprecedented temporal resolution of one minute, obtained by the SupraThermal Electron Proton sensor of the Energetic Particle Detector onboard Solar Orbiter at 0.85 au outside the He focusing cone. The instrument was not designed for pickup ion measurements, but provides unexpected advantages compared to conventional pickup ions instruments. Our results show that torus-shaped velocity distribution functions are variably on the one minute time scale. 
   \change{The measured pickup ion count rates indicate that these tori are not solely local structures, instead, these tori are populated by a large fraction of the total number of pickup ions injected in the past of the same solar wind stream and propagate to 0.85 au with at most weak pitch-angle scattering.} 
   }

   \keywords{Astroparticle physics --
                Scattering --
                Solar wind --
                ISM: general
               }

   \maketitle
   \nolinenumbers
\section{Introduction}\label{sec.introduction}
The heliosphere \citep{1961ApJ...134...20P} is surrounded by the local interstellar medium~\cite[LISM;][]{1987ARA&A..25..303C}. 
Due to the relative motion of the heliosphere against the LISM, a continuous stream of interstellar neutrals (ISNs) enters the heliosphere with a defined speed and direction.
Within the heliosphere, ISNs are subjected to gravitational focusing and ionization processes.
The latter depend on location and solar activity through photoionization, electron-impact ionization, and charge exchange \citep{2019ApJ...872...57S}.
Gravitational focusing leads to a focusing cone \citep{1968Natur.219..473F,1974ApJ...193..471W,1984A&A...134..171D,1995A&A...304..505M,2004A&A...426..845G,2004A&A...426..867L,2012JGRA..117.9106D} downwind of the ISN inflow direction, wherein the neutral He densities and consequently the $\mathrm{He}^{+}$ production rates are higher than outside the focusing cone.
Once an ISN is ionized, it is picked up by the solar wind plasma and is referred to as an interstellar pickup ion~\cite[PUI;][]{1985Natur.318..426M}.

PUIs are a fundamental constituent of the heliosphere and its interaction with the LISM \citep{1996JGR...10115523B,2022SSRv..218...18S}. 
They also serve as test particles to study fundamental plasma processes.
Within a few astronomical units (au) \change{of the Sun} interstellar pickup $\mathrm{He}^{+}$ is the most abundant interstellar PUI species \citep{2004JGRA..109.2104M,2013A&A...557A..50B,2023ApJ...950...98K}.

Interstellar PUIs are continuously injected into the solar wind and can be identified by their low charge states and their characteristic velocity distribution functions (VDFs) \citep{2022SSRv..218...28Z}.
When PUIs are injected, they are subject to the Lorentz force exerted by the convected-by local magnetic field and, thus, spiral around the magnetic field direction.
Thereby, their VDF initially forms a torus in velocity space, at a pitch angle (PA) that is defined by the local magnetic field,  solar wind conditions, and neutral velocity. 

In the dynamic solar wind, such a torus VDF is subjected to multiple processes, such as cooling, heating, PA scattering, or magnetic focusing, that modify its initial shape. Further, depending on their PA PUIs move relative to the solar wind bulk. 
Most theoretical descriptions of PUI VDFs assume rapid PA scattering to transform the initial torus VDF into an isotropic shell VDF~\citep{1976JGR....81.1247V,1986JGR....91.9965I,1988Ap&SS.144..487M,2025SSRv..221..108R}.
In the expanding solar wind, these shells are expected to be subject to cooling, i.e. collective energy loss.
Consequently the continuous injection would result in a filled sphere VDF.
For decades observations of PUI VDFs have been studied based on these assumptions \citep{2022SSRv..218...28Z,2025SSRv..221..100M}. 
However, there is ample evidence for anisotropies related to the local magnetic field conditions \citep{2002GeoRL..29.1612O,2015A&A...575A..97D,2020ApJ...897....6S,2025ApJ...981...35O}.

In the past the study of clear anisotropic torus signatures and the time scales of modifying processes have been limited by instrumental capabilities \citep{1985Natur.318..426M,1986AdSpR...6a.199M,2004A&A...426..845G,2025A&A...696A.115K,2020ApJ...897....6S,1999GeoRL..26.3181M,1998SSRv...86..127G,2015A&A...575A..97D,2002GeoRL..29.1612O,2025ApJ...981...35O}.
So far, the majority of PUI studies was based on reduced VDFs and averaged over time periods significantly longer than the typical fluctuations in the solar wind conditions. 

Despite the huge number of studies on He$^{+}$ PUIs, we found no He$^{+}$ number densities obtained from measurements. Mostly differential phase space densities are reported. \citet{2013JGRA..118.1389G} give time series of partial He$^{+}$ densities observed by the Fast Imaging Plasma Spectrometer onboard MESSENGER between 0.3 and 0.7 au and the Solar wind Ion Composition Spectrometer onboard the Advanced Composition Explorer at $\approx 1$ au.
During solar activity minimum and outside the focusing cone they find values that are in the range $10^{1}$ m$^{-3}$--$10^{2}$ m$^{-3}$. These are partial densities integrated over the spinning field of view (FOV) of the respective instrument without any assumption on the PUI VDFs. Taking the coverage of both instruments and the lower total ionization rates during solar activity minimum into account the total He$^{+}$ number densities can be expected to be up to about one order of magnitude higher for our period of interest and, thus, in the range $10^{2}$ m$^{-3}$--$10^{3}$ m$^{-3}$.  

The data and instrument are introduced in Sect. \ref{sec.DATA}.
For comparison, we conducted theoretical estimations on the torus VDF properties detected with different instrumental capabilities in Sect. \ref{sec.cts}.
A virtual detector is developed to simulate the instrument response to different VDFs, details are given in \change{Sect.~\ref{sec.signatures} and Appendix \ref{sec.vd}}.
The observations are then presented in Sect.~\ref{sec.results} and are analyzed with methods described in Appendix \ref{sec.totalcts} and \ref{sec.toruswidth}.
The observations are discussed in Sect.~\ref{sec.discussion}, followed by the conclusion section.

\section{Data and instrument}\label{sec.DATA}
Data used in this work and the flow parameters of ISNs applied here are introduced in Sect. \ref{sec.data}. The instrument that provides PUI measurements is introduced in Sect. \ref{sec.instrument}

\subsection{Data}
\label{sec.data}
For investigations of PUI VDFs, more than observations of the PUIs themselves are required.
Their physical frame of reference is determined by the solar wind plasma which carries the solar magnetic field away from the Sun.
In addition, electron data can provide valuable context on local ionization and thereby PUI production rates. 
Therefore, we use data from four instruments on Solar Orbiter \citep{2020A&A...642A...1M} that are publicly available from the Solar Orbiter Archive\footnote{ \url{https://soar.esac.esa.int/soar/}~(Retrieved June, 1, 2025).}.
With the exception of the spacecraft location and the Eqs. \ref{eq.prodrate} -- \ref{eq.int2}, for all measurements and derived quantities in this study we use the common coordinate system of Solar Orbiter, the spacecraft reference frame (SRF), which is illustrated in Fig.~\ref{fig.1}a. 
In the SRF, the $x$-axis points toward the Sun.

The PUI observations we report in this study are obtained from the $1$-second level 1 data of the Suprathermal Electron Proton (STEP) instrument of the Energetic Particle Detector \cite[EPD;][]{2020A&A...642A...7R}.
The STEP instrument and its data are further described in \change{Sect.~\ref{sec.instrument} and Sect.~\ref{sec.signatures}}.
Magnetic field vectors $\bm B$ are obtained from the $8$ Hz resolution data (level 2) of the Magnetometer \cite[MAG;][]{2020A&A...642A...9H}.
Solar-wind proton ground calculated moments (level 2), density $n_\mathrm{sw}$, velocity $\bm{v}_\mathrm{sw}$, and temperature $T_\mathrm{sw}$, and energy spectra (level 1) come from the Proton-Alpha Sensor of the Solar Wind Analyser \cite[SWA-PAS;][]{2020A&A...642A..16O} at $4$-second time resolution. 
Solar-wind electron energy spectra (level 1) are taken from the Electron Analyser System of the Solar Wind Analyser \cite[SWA-EAS;][]{2020A&A...642A..16O} at $10$-second time resolution.

Further, we use spacecraft location and attitude information provided by the ESA SPICE Service \citep{1996P&SS...44...65A,2018P&SS..150....9A}. 
On the start of November 5, 2021, Solar Orbiter was at a radial distance $R=0.85~\mathrm{au}$, longitude $\lambda = 41.9^\circ$, and latitude $\beta =-2.0^\circ$ in ecliptic coordinates. 
The spacecraft was not in the focusing cone of PUIs \citep{1968Natur.219..473F,1995A&A...304..505M,2004A&A...426..845G,2012JGRA..117.9106D}, \change{whose center} is $\lambda = 75.3^\circ$, and $\beta = -5.1^\circ$. 
For this location and the spacecraft orientation we calculated the expected local neutral $\mathrm{He}$ velocity, which is used as the local injection velocity, $\bm v_{\mathrm{inj}}$, of PUIs.
Therein, we assumed an ISN inflow velocity of $v_{\infty}=25.4$ km/s, and an inflow direction in ecliptic coordinates of $\lambda = 75.3^\circ$, and $\beta = -5.1^\circ$.
These parameters are an estimate based on the different values that are given in the literature \citep{2004A&A...426..835W,2012JGRA..117.9106D,2012Sci...336.1291M,2012ApJS..198...12B,2012ApJS..198...11M,2014A&A...569A...8B,2015ApJ...801...62W,2015ApJ...801...28M,2015ApJ...804...42L,2015ApJS..220...28B,2018A&A...611A..61T,2023ApJ...953..107S}.
The temperature of ISNs \change{in the LISM} and the gravitational deformation of the thermalized neutral $\mathrm{He}$ distribution \citep{2012AIPC.1436..233M}  are neglected in this study. Only the gravitational focusing and acceleration of the mean of the distribution is considered. 
With these values we find \change{injection} velocities \change{of} $[0.94\pm0.21,-21.44\pm0.01,-2.70\pm0.12] $ km/s, in the SRF on November 5. All times are given in universal time (UT).

\subsection{STEP's principle of operation}\label{sec.instrument}
\begin{figure*}
    \centering
    \includegraphics[width=17cm]{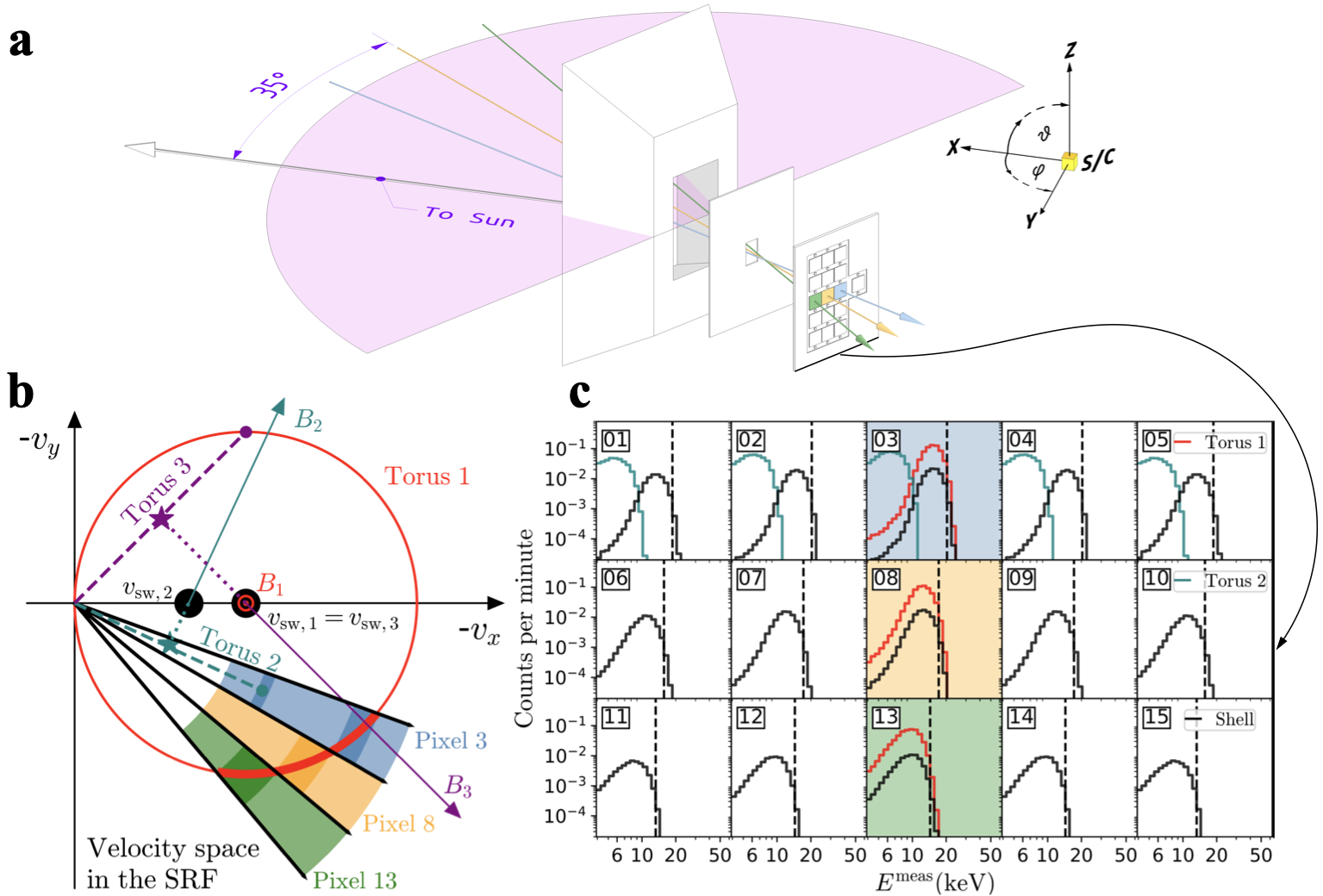}
    \vspace{-0.15cm}
    \caption{Sketch of STEP/IC, three VDFs with respect to STEP's FOV in velocity space, and artificial measurements.
    \textbf{a,} Sketch of STEP/IC as a pinhole camera with three example trajectories. The SRF coordinate system is illustrated on the top right.
     \textbf{b,} A sketch of \change{a 2D cut of the} three torus-shaped VDFs in velocity space in the SRF, and FOV of three pixels of STEP/IC. 
     \change{Torus 1 (red), 2 (blue) and 3 (purple), are defined by their different solar wind conditions, i.e. the respective solar wind bulk velocities indicated by black filled circles define the torus center, $\bm{v}_{\mathrm{sw}, 1} = \bm{v}_{\mathrm{sw}, 3}= [600,0,0]~\mathrm{km/s}$ and $\bm{v}_{\mathrm{sw}, 2} = [400, 0, 0]~ \mathrm{km/s}$, and the respective magnetic field vectors with elevation angles $\vartheta_1=90^\circ$, $\vartheta_2=\vartheta_3=0^\circ$, and azimuthal angles $\varphi_1$~(degenerate), $\varphi_2=245^\circ$, $\varphi_3=135^\circ$. Each magnetic field vector is attached to the corresponding solar wind bulk velocity. $\bm{B_1}$ points orthogonal to the $v_x$-$v_y$ plane and is  indicated with a red dot on top of the marker for the  corresponding  $\bm{v}_{\mathrm{sw}, 1}$. The solid red line of Torus 1 indicates that Torus 1 is located in the $v_x$-$v_y$ plane, and its intersection with STEP's FOV is marked with a thicker red line. 
     For Torus 2 and 3, the projections of the torus to the $v_x$-$v_y$ plane are depicted as dashed lines, their guiding centers are marked as stars, and the intersections (outside the origin) with the shown plane are marked with colored dots.}
     The neutral He velocity is neglected in this sketch but not in the calculations.
    Color shaded regions indicate STEP energy ranges.
    \textbf{c,} Artificial energy spectra that correspond to \change{Torus 1 and Torus 2 in \textbf{b} and an artificial shell VDF with the same radius as Torus 1,} obtained with the virtual STEP detector described in Appendix.~\ref{sec.vd}. 
    \change{For all  artificial VDFs the same input number density was used. }
    \change{Dashed lines are cutoff energies of the shell VDF on the individual pixels, i.e. at 10$\%$ of the respective maximum.}
    }
    \label{fig.1}
\end{figure*}
The PUI observations presented here were obtained from STEP, which was designed to measure ions and electrons at suprathermal energies. 
It consists of two co-aligned sensor heads with a parallel fixed FOV of $\approx 28^\circ\times54^\circ$ centered \change{on} the expected average Parker angle, see Fig.~\ref{fig.1}a. 
One of the sensor heads contains a permanent magnet to deflect electrons (magnet channel, MC), the other does not (integral channel, IC). 

The principle of operation is depicted in Fig.~\ref{fig.1}a. Each sensor head operates as a particle  pinhole camera.
Particles from all directions within the FOV enter through a pinhole and are then detected by a $3\times5$ pixel solid-state detector \citep{2020A&A...642A...7R}.
In Fig.~\ref{fig.1}a, the trajectories of three example particles are shown in three colors that hit three color-shaded pixels, respectively.
Each pixel views a certain direction and \change{each of the 15 pixels provide a $\approx10^\circ \times 15^\circ$ FOV.}

Within the solid-state detector, particles lose energy, and the ionizing energy losses in the active volume of the detector are measured.
For the main data product used in this study, the recorded energy loss of each incident particle is sorted into 32 logarithmically spaced energy bins in the range $3.92$~keV - $62.72$~keV.
Each second for each pixel, a histogram with the number of counts per energy bin is collected.
In this energy range, particles are stopped in the detector and the recorded energy is a measure for their initial kinetic energy. 
However, due to energy losses in dead layers and  non-ionizing energy losses in the active volume the fraction of the kinetic energy that is measured in the active volume of the solid state detectors is of stochastic nature and depends on the particle species, which can not be determined \change{directly} by STEP.
\change{The conversion from measured energy to primary particle energy requires a species dependent calibration (see Appendix \ref{sec.vd}).
Further, the 15 pixels are subject to individual aging effects making the calibration time dependent.
This calibration of this aging effect is ongoing (see Appendix \ref{sec.aging}) and is therefore not corrected for in the following.} 

For periods of enhanced suprathermal particle fluxes, it is typically assumed that protons and electrons are the \change{dominant} species.
Under this assumption, the MC measures only protons, the IC protons and electrons, and electron measurements can be obtained by the difference between IC and MC.
\change{These measurements can then be converted into proton and electron fluxes using the respective instrumental responses.}
\change{Without the assumption of protons being the dominant ion species, i.e. strong contributions of helium (or other ions) to the measurement, the MC mixes contributions of different ion species.}
\change{While for protons the angular response of the MC is designed to be similar to the angular response of the IC, for helium and other heavier ions the angular responses of MC and IC differ more strongly than for protons}.
\change{This difference is not well characterized and introduces a species dependent uncertainty to the directional information provided by STEP's pixels.} 
\change{Therefore, a}fter verifying that both IC and MC show similar results\change{, i.e. no significant electron contribution is observed} for our time period of interest, in this study we only consider the IC. \change{Any possible remaining minor electron contributions are treated as part of the background subtraction, see Appendix \ref{sec.totalcts}.}

The calibrated energy channels provided with the main data product are for protons with kinetic energies in the range $5.73$ keV - $68.00$ keV. 
Although $\mathrm{He}$ PUIs have kinetic energies in this range, this calibration cannot be used directly for $\mathrm{He^+}$ PUIs.
\change{In this study we calculated and utilized $\mathrm{He^+}$ responses that have been derived similar to the proton responses (see Appendix \ref{sec.vd}).}

\section{PUI torus VDF properties}\label{sec.cts}
PUIs are injected continuously into the solar wind plasma.
As a consequence, the PUI VDFs observed at a certain location have been injected over an extended time period over an extended spatial volume which contribute to a total He$^+$ number density, $\niontot$. In this section, we discuss how PUI number densities can be derived, and how injection time scales and injection path lengths can be constrained from STEP observations.

Information about remote plasma conditions is embedded in the PUI VDFs.
In turn, the observed VDFs can not uniquely be interpreted in the context of the locally measured plasma conditions, such as solar wind speed and magnetic field direction.
If we observe torus signatures that are clearly related to the local solar wind velocity and magnetic field direction and if we assume that also the local PUI generation rate is known, corresponding temporal and spatial scales can be estimated on which the injection conditions would need to have been stable to explain the locally observed number density, $\niontorus$, in a torus VDF.

\subsection{Densities}
In this section, we discuss the instrumental aspect of the determination of torus densities $\niontorus$.
PUI instruments integrate the differential number density, $\delta n(\bm{v})$, of ions over their FOV, i.e. over a volume in velocity space and time, which corresponds to a volume in position space.
In a general way the expected number of counts, $\hat{C}$, can be expressed by
\begin{equation}
\label{eq.intinstr}
\hat{C} = \int\limits_{-c}^{c}\int\limits_{-c}^{c}\int\limits_{-c}^{c} \delta n(\bm{v}) \cdot v \cdot \taum \cdot \zeta(\bm{v}) \cdot A(\bm{v}) \cdot \eta(\bm{v}) ~  dv_x dv_y dv_z\enspace
\end{equation}
Therein, $\bm{v}$ is the velocity vector of ions in the instrumental frame of reference, $v$ its absolute value, and $c$ is the speed of light.
$\taum$ is the measuring time interval and $\zeta(\bm{v})$ is the duty cycle, i.e. the fraction of $\taum$ wherein ions with $\bm{v}$ can be measured.
$A(\bm{v})$ is the active area of the detector as described by the directional response, and $\eta(\bm{v})$ is the efficiency to detect an ion. 
Combining all instrumental properties we obtain differential responses, $\delta R(\bm{v})$,  
\begin{equation}
\label{eq.diffresp}
\delta R(\bm{v}) = v \cdot \taum \cdot \zeta(\bm{v}) \cdot A(\bm{v}) \cdot \eta(\bm{v})
\end{equation}
and can rewrite Eq.~\ref{eq.intinstr} as
\begin{equation}
\label{eq.intinstr2}
\hat{C} = \int\limits_{-c}^{c}\int\limits_{-c}^{c}\int\limits_{-c}^{c} \delta n(\bm{v}) \cdot \delta R(\bm{v})~  dv_x dv_y dv_z\enspace .
\end{equation}
All quantities in Eq.~\ref{eq.intinstr} depend on the ion species and can be evaluated either for the whole instrument or e.g. for individual energy bins and viewing directions.
Detailed information about $\delta n(\bm{v})$ is lost during the measurement process and only information about the integrated density over the scanned volumes can be obtained from Eq.~\ref{eq.intinstr2}.
It is not even possible to derive the corresponding integrated density without further assumptions about $\delta n(\bm{v})$.  
Here commonly the assumption of constant $\delta n$ and $\delta R$ over the scanned spatial and velocity space volumina is made.
Under this assumption the differential responses $\delta R(\bm{v})$ can be approximated by averaged effective responses $\delta R_\mathrm{eff}$ and corresponding effective velocity $v_\mathrm{eff}$, effective duty cycle $\zeta_\mathrm{eff}$, effective area $A_\mathrm{eff}$, and effective efficiency $\eta_\mathrm{eff}$
\begin{equation}
    \delta R_\mathrm{eff} = v_\mathrm{eff} \cdot \taum \cdot \zeta_\mathrm{eff} \cdot A_\mathrm{eff} \cdot \eta_\mathrm{eff} \cdot H(\bm{v})\enspace ,
\end{equation}
with $H(\bm{v}) = 1$ if $\bm{v}$ is inside and $H(\bm{v}) = 0$ if $\bm{v}$ is outside the FOV.
\begin{eqnarray}
\label{eq.avgcts}
\hat{C} & = & \delta n \cdot \delta R_\mathrm{eff} \cdot \int\limits_{-c}^{c}\int\limits_{-c}^{c}\int\limits_{-c}^{c} H(\bm{v}) ~ dv_x dv_y dv_z\\
\label{eq.intresp1}
 & = & \delta n \cdot V_{\mathrm{vspace}} \cdot \delta R_\mathrm{eff}\\
 & = & n^{\mathrm{FOV}} \cdot \delta R_\mathrm{eff}
\end{eqnarray}
with the velocity space volume $V_{\mathrm{vspace}}$. Then the FOV integrated density, $n^{\mathrm{FOV}}$, can be calculated by
\begin{equation}
    n^{\mathrm{FOV}} = \frac{\hat{C}}{\delta R_\mathrm{eff}} .
\end{equation}

Due to their limited FOV, PUI instruments do not measure the total He$^+$ number density directly, but a partial number density $\nionpart$ which can be obtained from the measured number of counts, $\cts$, by
\begin{equation}
\label{eq.cts1}   \nionpart = \frac{\cts}{\delta R_{\mathrm{eff}}}\enspace.    
\end{equation}
Therein, $v_{\mathrm{\mathrm{He}^{+}, eff}}$ is taken as the velocity of He$^{+}$ in the SRF in the center of the FOV.
The partial densities can only be scaled to total densities if assumptions about the whole VDF are made.
If we observe signatures of torus VDFs the number of counts that are associated with the torus signature, $\ctstorus$, can be used to calculate the number density of the whole torus as the fraction of the torus $\omega$ that is covered by the FOV of the instrument 
\begin{equation}
\label{eq.ntorus}
\nionpart= \omega \cdot \niontorus\enspace .
\end{equation}
As an example for $\omega$ one can consider the configuration of Torus 1 (red circle) in Fig.~\ref{fig.1}b\footnote{Fig.~\ref{fig.1} b is discussed in detail in Sect.~\ref{sec.signatures}.}.
For the magnetic field direction of Torus 1 the complete torus lies in the $v_x,v_y$-plane and $\omega$ is the fraction of the length of Torus 1 that crosses the FOV of STEPs pixels 3, 8, and 13 (bold red in Fig.~\ref{fig.1}b) to the total length of Torus 1.
With Eqs.~\ref{eq.cts1} and \ref{eq.ntorus} we obtain
\begin{equation}
\label{eq.ntorus2}   \niontorus = \frac{\ctstorus}{\delta R_{\mathrm{eff}} \cdot \omega} \enspace.    
\end{equation}
\subsection{Injection time scale and injection path length scale}\label{sec.T_inj.L_inj}
PUIs are continuously injected into and are convected away from the  Sun with the expanding solar wind plasma.
In addition the PUI VDF is continuously evolving in the ambient plasma.
Thus, information about the whole history of the evolution of PUIs from the Sun to the point of observation is embedded in the in-situ observed PUI VDF.
The complex interplay of various processes that can alter the PA and/ or the energy of PUIs as well as the unknown injection conditions in the history of the observed PUI VDFs complicates the interpretation and is always based on several assumption. 
Here we make the assumption that an observed torus that is related to the in-situ injection conditions, i.e. magnetic field direction and solar wind velocity, is a result of stable injection conditions over a possibly extended period, and possibly extended injection path-length, respectively.
In a torus, individual PUIs gyrate about their guiding center. For the examples in Fig.~\ref{fig.1}b, the speed of the guiding center in the SRF is always below the solar wind speed. This generalizes to other PUI torus VDFs in the solar wind. The only exception are PUI VDFs with a $90^\circ$ PA. Only in this case, the guiding center coincides with the solar wind bulk. As a result, after injection the guiding center typically lags behind the solar wind bulk and moves relative to the structures convected with the solar wind bulk. Here, we assume a convection with the solar wind bulk velocity.

These assumptions implicitly neglect any PA scattering, magnetic focusing, or energy changes of PUIs.
With these assumptions $\niontorus$ is the result of the combination of continuous injection of PUIs and a thinning due to the expansion of the solar wind.
We first estimate the production rate of locally produced He$^+$, $\dnion(r)$ at a radial distance $r$ as 
\begin{equation}
\label{eq.prodrate}
\dnion(r) = \nn(r)\cdot \I(r)\enspace ,
\end{equation}
with position-dependent local neutral He density $\nn(r)$ and local total ionization rate $\I(r)$. 
The produced He$^+$ number density, $\nionprod$, resulting from all He$^+$ PUIs injected into a solar wind stream between a time $t_0$ and $t$ with corresponding positions $r(t_0)=R_0$ and $r(t)=R$ is then obtained by integration
\begin{equation}\label{eq.int}
    \nionprod(t-t_0)=\int\limits_{t_0}^t \dnionp \frac{r(t^\prime)^{2}}{R^{2}} \mathrm{d} t^\prime \enspace .
\end{equation}
The term $\frac{r(t^\prime)^{2}}{R^{2}}$ accounts for the expansion of the solar wind plasma which thins out PUIs injected closer to the Sun.
The time dependent position $r(t)$ is obtained by $r(t) = \vc \cdot t$ 
with the convection speed $\vc$.
We now assume the local total ionization rate to scale quadratically with the distance to the Sun, and $\I(r)$ is given relative to the local total ionization rate $\Ir{R}$ at a reference position $R$ as
\begin{equation}
\label{eq.ionrate}
    \I{}(r) = \frac{R^2}{r^2} \cdot \Ir{R}\enspace. 
\end{equation}

Further, we now assume the neutral helium density to be constant $\nn(r) = \nn$.
Inserting both in the integral (Eq. \ref{eq.int}) and substituting $\mathrm{d}t=\mathrm{d}r \cdot \vc^{-1}$ with a constant convection speed  $\vc$ yields
\begin{eqnarray}
    \nionprod(R-R_0) & = &\frac{1}{\vc} \int\limits_{R_0}^R \nn \cdot \frac{R^2}{r^2} \cdot \Ir{R}  \cdot \frac{r^2}{R^2} \quad\mathrm{d} r \label{eq.intexp}\\
          & = & \frac{\Ir{R} \cdot \nn}{\vc} \int\limits_{R_0}^R \mathrm{d} r \\
          & = & \Ir{R} \cdot \nn\frac{(R-R_0)}{\vc} \enspace .\label{eq.int2}
\end{eqnarray}
We now define a proxy for the injection time scale, $\tage = \frac{R-R_0}{\vc} = t-t_0$.
The He$^+$ number density produced within $\tage$ is estimated as
\begin{equation}
\label{eq.puidens}
   \nionprod(\tage) = \Ir{R} \cdot \nn \cdot \tage \enspace \enspace,
\end{equation}
with a corresponding proxy for the injection path length $\linj= \tage \cdot \vc = R-R_0$.
For the computation of $\linj$, we assume $\vc=\vsw$.
Using Eqs. \ref{eq.ntorus2},\ref{eq.puidens}, we can calculate $\tage$ and $\linj$ from $\niontorus$ and $v_\mathrm{sw}$, by
\begin{eqnarray}
    \label{eq.tage}
    \tage(\ctstorus) & = & \frac{\ctstorus}{\delta R_{\mathrm{eff}} \cdot \omega \cdot \Ir{R} \cdot \nn},\\
    \label{eq.linj}
    \linj(\ctstorus) & = & \frac{\ctstorus \cdot \vsw}{\delta R_{\mathrm{eff}} \cdot \omega \cdot \Ir{R} \cdot \nn}
\end{eqnarray}

We consider $\tage$ and $\linj$ as proxies  for the unknown real injection time and length scales since both rely on the assumptions that were made to derive Eq. \ref{eq.int2}.
Thus, both proxies are based in particular on the assumption of a constant neutral helium density which is only justified on small spatial and temporal scales, because no depletion of neutral helium with decreasing radius is taken into account.
Further our proxies assume that the thinning by the solar wind expansion is canceled by the decreasing ionization rates, which effectively results in constant production rates, $\dnion(r) =\Ir{R} \cdot \nn$.
If the derived injection time proxy is short, that is, in the order of seconds or minutes, the above assumptions are likely to be justified and the observed torus feature is indeed the result of only local injection.
If the injection time scale proxy is long, that is, in the order of hours or even in the order of the travel time of the solar wind from the Sun to the spacecraft, the assumption of constant production rates is likely to be violated. In this case, the injection time and injection length proxies from Eqs.~\ref{eq.tage} and \ref{eq.linj} would imply strong contributions from non local injection. 

\subsection{One count $\tage$ and $\linj$}\label{sec.1count}
In preparation for the discussion in Sects.~\ref{sec.results} and \ref{sec.discussion}, we employ Eqs. \ref{eq.tage} and \ref{eq.linj}, to derive injection time and path length proxies for a measurement of a single count $\tage(1)$ and $\linj(1)$ for Torus 1 in Fig.~\ref{fig.1}b and for three instruments.
The resulting values  are  summarized in Table~\ref{tab.tage}.
These proxies serve to estimate the temporal and spatial scales on which PUIs would need to be injected under the above discussed assumptions to result in a torus density $\niontorus$ that corresponds to an expectation of a single measured count for  different instruments.
In the following the input parameters for the calculations are described.
Since no reference $\nn$ measurements at the same time and place are available, we use the interstellar value, $\nninf=1.5\cdot10^{4}~\mathrm{m}^{-3}$~\citep{2019ApJ...882...60B,2025ApJ...991..122I}.  
No depletion of neutral He throughout the heliosphere due to ionization is taken into account here, and thereby this value deliberately overestimates the local He density outside the focusing cone. 
Solar activity dependent local total ionization rates $I_{\mathrm{He}}(1 ~\mathrm{au})$ are derived in \citet{2015ApJS..220...27S}.
For the period of interest we find values well below $\mathrm{10}^{-7}~\mathrm{s}^{-1}$.
Again, as an overestimation we assume a rate of $\mathrm{10}^{-7}~\mathrm{s}^{-1}$ at 1 au, resulting in $I_{\mathrm{He}}(0.85~\mathrm{au}) = 1.4  \cdot \mathrm{10}^{-7}~\mathrm{s}^{-1}$ with Eq. \ref{eq.ionrate}. 
With these values we obtain from Eq.~\ref{eq.prodrate} a local production rate $\dnion(0.85~\mathrm{au}) =2.1 \cdot 10^{-3}~\mathrm{m^{-3}s^{-1}}$ which  is utilized in all following calculations.
Similarly, in the following we consider a fixed measuring time interval $\taum = 60$s. 

For STEP we consider $\zeta_\mathrm{eff} = 1$ and $\eta_\mathrm{eff} = 1$, i.e. all energies and directions are measured 100\% of the time and He$^{+}$ is measured with $100\%$ efficiency.
The FOV averaged effective area of STEP is $A\mathrm{_{eff}} = 2 \cdot 10^{-6} ~\mathrm{m}^{-2}$ (compare Fig.~\ref{fig.energy.response}).
Further we assume the situation of Torus 1 shown in Fig.~\ref{fig.1}b, with $v_{\mathrm{He}^{+}, \mathrm{eff}} = 9.83\cdot10^{5}~\mathrm{m/s}$ (He$^+$ velocity in the center of pixel 8) and the fraction $\omega= 0.156$ which corresponds to the three dark shaded areas in Fig.~\ref{fig.1}b.

For a baseline comparison to other active instruments, we also calculated expected $\tage(1)$ and $\linj(1)$ for the main channel of the solar wind section of the PLasma And SupraThermal Ion Composition \citep[PLASTIC;][]{Galvin2008} instrument of the Solar TErestrial RElations Observatory-Ahead (STEREO-A) which has been used in several PUI studies.
PLASTIC scans different parts of the velocity space consecutively. Therein, PLASTIC scans its FOV in 128 energy-per-charge steps and 32 deflection steps, i.e. $\zeta_{\mathrm{eff}}=1/128 \cdot 1/32$, within $\taum = 60 \mathrm{s}$, and a He$^+$ efficiency $\eta_{\mathrm{eff}}\lesssim 0.5$ \citep{Galvin2008}. PLASTIC has an effective area $A_{\mathrm{eff}}= 10^{-5}\mathrm{m}^{2}$, and an intersection of the Torus 1 with its FOV $\omega=0.25$.  The He$^{+}$ velocity at the center of the FOV $v_{\mathrm{\mathrm{He}^{+}, \mathrm{eff}}} = 1.2\cdot10^{6}~\mathrm{m/s}$.

As an additional reference, the expected $\tage(1)$ an $\linj(1)$ for PAS for the same Torus 1 example is included in Table~\ref{tab.tage}. The operation principle of PAS is similar to PLASTIC's but lacks a time-of-flight section. With a duty cycle $\zeta_\mathrm{eff} = 1/96 \cdot 1/9$, an intersection of the PAS FOV with the torus $\omega = 0.37$, an average effective area $A_\mathrm{eff}= 4 \cdot 10^{-8} \mathrm{m}^{2}$ \citep{2025A&A...702A.135B}, a He$^+$ velocity at the center of the FOV $v_\mathrm{He^+, eff} = 1.185 \cdot  10^6$~m/s, an efficiency $\eta_\mathrm{eff} = 1$, and an integrated measurement time interval $\taum = 60~\mathrm{s}$. We note, that for this torus configuration, $v_\mathrm{He^+}$ in the center of PAS's FOV is already higher then the highest He$^+$ velocity ($\approx 9.43 \cdot 10^5$  m/s) PAS can resolve. 

In addition, Table \ref{tab.tage} lists the corresponding one-count proxies, $\tage_{\mathrm{VDSTEP}}(1)$ and $\linj_{\mathrm{VDSTEP}}(1)$, for Torus 1 derived based on the virtual STEP detector described in Appendix \ref{sec.vd}.
\begin{table}
\centering
\caption{Different estimates for $\tage(1)$ and for $\linj(1)$  of an example VDF (Torus 1 in Fig. \ref{fig.1}b).}
\label{tab.tage}
\begin{tabular}{cccc}
\hline\hline\noalign{\smallskip}

$\tage_{\mathrm{VDSTEP}}(1)$  & $\tage_{\mathrm{STEP}}(1)$    & $\tage_{\mathrm{PLASTIC}}(1)$ & $\tage_{\mathrm{PAS}}(1)$\\
(min) & (min) & (min) & (min)\\
\midrule
$0.42$ & $0.55$  &  $333.34$ & $5555.55$ \\
\bottomrule
\hline\hline\noalign{\smallskip}

$\linj_{\mathrm{VDSTEP}}(1)$  & $\linj_{\mathrm{STEP}}(1)$    & $\linj_{\mathrm{PLASTIC}}(1)$ & $\linj_{\mathrm{PAS}}(1)$\\
(au) & (au) & (au) & (au)\\
\midrule
$1.0 \cdot 10^{-4}$ & $1.3 \cdot 10^{-4}$  &  $8.0 \cdot 10^{-2}$ & $1.3$  \\
\bottomrule
\end{tabular}
\tablefoot{Values with subindices STEP, PLASTIC and PAS are calculated for the respective instruments from Eqs.~\ref{eq.tage} and \ref{eq.linj}. Subindex VDSTEP indicates that the values have been derived from a virtual detector of STEP with full responses and not by the simplified equations.}
\end{table}

\section{\change{Methods for identifying He$^+$ tori with STEP}} \label{sec.signatures}
Here, we discuss what signatures of He$^\mathrm{+}$ PUIs are expected in STEP and how these signatures can be utilized to identify He$^\mathrm{+}$ PUIs in STEP even though STEP is not designed for species identification. We illustrate the expected signatures with the help of a virtual STEP detector which is described in detail in Appendix~\ref{sec.vd}.
\subsection{Signatures of He$^+$ torus and shell VDFs in STEP and He$^+$ identification}\label{sec.4.1}
As outlined in Sect.~\ref{sec.introduction}, initially PUIs form a torus VDF in the solar wind frame of reference.
\change{Further, it is expected that PA scattering} then transforms these torus VDFs into shell VDFs. These VDFs are then subject to cooling or heating and ultimately lead to filled sphere VDFs.
Here, we focus on the two cases that are expected to describe the VDFs of recently created PUIs, namely a torus VDF and a fully scattered shell VDF. 

Figure \ref{fig.1}b illustrates the velocity space coverage of STEP pixels 3, 8, and \change{13} (color shaded regions) in the $v_\mathrm{x}$--$v_\mathrm{y}$ plane in the SRF.
In addition, three illustrative torus-shaped VDFs for three different magnetic field configurations and solar wind velocities are sketched.
A small non-zero $\bm v_\mathrm{inj}$ of the neutral particles is omitted in the sketch, but taken into account in our analysis (see Sects.~\ref{sec.data}, \change{\ref{sec.signatures}} and \ref{sec.results}).
The figure illustrates how the solar wind \change{velocity} and magnetic field direction determine the position and visibility of a torus VDF in STEP's FOV. 

Torus 1 (red) is detected by all STEP pixels in the shown $v_\mathrm{x}$--$v_\mathrm{y}$ plane but Torus 2 (teal) would only be visible in pixel 3 \change{(and pixels 1, 2, 4, and 5 outside the shown $v_\mathrm{x}$--$v_\mathrm{y}$ plane)}.
Torus 3 (pink) is not detected at all within STEP's FOV. 
Although the illustrated tori are narrow, that is, approximately monoenergetic in the solar wind frame of reference, in STEP Torus 1 and Torus 2 appear smeared out in energy on the respective pixels in the SRF. This broadening of the torus signature is indicated with the darker shaded areas in Fig.~\ref{fig.1}b. 
In Figs.~\ref{fig.overview} and \ref{fig.90min} the same energy uncertainty for the expected primary energy of He$^{+}$ PUIs is marked with a white-shaded band.  

Figure \ref{fig.1}c shows the resulting virtual measurements of the virtual STEP detector described in Appendix~\ref{sec.vd} for the same tori examples in the same respective colors as in Fig.~\ref{fig.1}b.
As the input for the virtual detector, for comparability  each torus VDF was filled with the same number density \change{derived from Eq.~\ref{eq.prodrate} for 1 minute of PUI production}, that is, the same number density in position space over different volumina in velocity space, and for numerical reasons we assigned a finite full width of $20$ km/s.
As expected, Torus 1 is observed only by pixels 3, 8, and \change{13}, Torus 2 is detected by pixels 1 to 5, and Torus 3 is not detected by any STEP pixel.
The virtual measurements show a clear and distinct cut-off towards higher energies on the individual pixels.
Therein, as expected from Fig.~\ref{fig.1}b, pixel 3 observes the torus signature at higher energies than all other pixels.
On all pixels the signatures are smeared out down to the lowest observed energies.
The expected geometrical broadening can also be seen in Fig.~\ref{fig.1}b (darker-color shaded regions) and is further increased by the energy responses shown in \change{the Appendix in} Fig.~\ref{fig.energy.response}. 
From Fig.~\ref{fig.1}c we also see that the expected cut-off and broadening strongly depend on the solar wind velocity, as Torus 2 is seen at lower energies than Torus 1.

In addition, the virtual measurement of a shell VDF for the same solar wind speed, number density, and width as Torus 1 is shown in black in Fig. \ref{fig.1}c.
The shapes of the shell signatures are similar to that of the torus signatures on the pixels that also observed Torus 1, but the respective maximum count rates per pixel are about one order of magnitude smaller.
In return the shell signatures are observed on all 15 pixels of STEP. 
\change{In addition, the  cutoff energies of the shell VDF (dashed lines) show a similar pattern as for the torus examples discussed above.}

\change{This pattern of the  cutoff energies of a torus or shell VDF in STEP depends on the velocity of these particles and the position of the center of the corresponding torus or shell in velocity space. Since different particle species, e.g. He$^\mathrm{+}$ and protons, with the same energy have different velocities, this allows for the special case of torus or shell VDFs a species identification with STEP.  This is discussed in Appendix \ref{sec.identification} and illustrated for He$^\mathrm{+}$ and for protons in Fig.~\ref{fig.sketch.H.ring}. }

Further, we use the virtually measured shell VDF to approximate the geometrical torus broadening described above \change{and the additional broadening due to the instrumental response functions}. To this end, \change{similarly to the dashed lines marking the cutoff in Fig.~\ref{fig.1}c} we track the energies where the shell signatures falls to 10\% of their respective maxima. These energies (calculated using the measured solar wind velocities) are marked with pairs of black lines in Fig.~\ref{fig.overview} and Fig.~\ref{fig.90min}, and with pairs of dashed colored lines in Fig.~\ref{fig.1minute}.

Based on these considerations, we expect STEP to be able to distinguish between shell-like and torus-like VDFs \change{and for such VDFs distinguish between He$^+$ and protons}.
A shell VDF would be seen on all pixels and would be independent of the magnetic field direction.
Whereas, under the otherwise same solar wind conditions a (narrow) torus VDF will only be observed by a few STEP pixels.
\change{The characteristic cutoff pattern supports the species identification.}

\subsection{\change{Criteria for expected torus intersections with STEP's pixels}}\label{sec.4.2}
In general, the possible positions of a torus of freshly injected PUIs is restricted to a surface of a sphere in velocity space which is centered at the solar wind bulk. 
This is the same surface that would be fully populated by a fully scattered but not yet cooled or heated shell VDF. The radius of this shell is constructed from the solar wind velocity $\bm v_\mathrm{sw}$ and the neutral He velocity $\bm v_\mathrm{inj}$. 
For a certain magnetic field direction, the position on this shell is given by its PA. 
Consequently, both the orientation of a PUI torus VDF and the central viewing direction of each pixel can be expressed as a PA, $\alpha_{\mathrm{torus}}$ and $\alpha_{\mathrm{pixel}}$, respectively.

In the following, we derive two criteria to characterize if a torus signature of freshly injected PUIs is expected on individual STEP pixels. 
For the first criterion, we compare $\alpha_{\mathrm{torus}}$ and $\alpha_{\mathrm{pixel}}$
\begin{equation}
    \dpa=\alpha_{\mathrm{pixel}}-\alpha_{\mathrm{torus}}.
\end{equation}
Thus, the respective PA difference\change{,} $\dpa$\change{,} expresses our expectation of whether a fresh, i.e. injected under the local solar wind conditions, PUI torus VDF is detectable on each individual STEP pixel. If we define the FOV of a STEP pixel as the area in which the angular response is above $5\%$ of its maximum, then a FOV of $\approx 10^\circ \times 15^\circ$ appears under approximately twice that angle in the solar wind frame of reference for conditions similar to Torus 1. 
Therefore, STEP can detect a torus VDF only on pixels with $|\dpa| < 15^\circ$. This criterion is well suited to provide an overview of how close to the pixels' viewing direction the currently expected PUI torus is expected.

However for a tighter estimate, depending on the torus geometry, different pixels would require different thresholds. Thus, we augment $\dpa$ with a second pixel-wise criterion that also takes the variability of the magnetic field direction into account. 
\change{Similar to $\omega$ as defined in Sect.~\ref{sec.cts},} for a given torus \change{(no angular width)}, $\omega_\mathrm{pixel}$ refers to the fraction of this torus that intersects with the pixel's FOV. For each magnetic field direction in $8$ Hz resolution, we calculate the respective torus and then compute for each pixel the average $\widetilde{\omega}_{\mathrm{pixel}}$ over the time scale of interest, which is one minute in this study. 
If $\widetilde{\omega}_{\mathrm{pixel}}>10^{-3}$ we consider this pixel to be hit during this minute, i.e. on average at least one thousandth of the torus' circumferential length intersected with the pixel's FOV.
This is indicated in Fig.~\ref{fig.1minute}a with color-coded outlines and with a \texttimes-symbol in Fig.~\ref{fig.1minute}b. Pixels that should not detect a torus signature of freshly injected He$^+$ PUIs are considered in the background estimation described in Appendix~\ref{sec.totalcts}.

Since STEP's FOV is oriented around the expected average Parker angle, magnetic field configurations that match Torus 1 in Fig.~\ref{fig.1}b are unusual under typical solar wind conditions and most frequently occur either in highly Alfv{\'e}nic coronal hole wind or associated with interplanetary coronal mass ejections (ICMEs).

\section{STEP observations}\label{sec.results}
\begin{figure*}[ht]
    \centering
    \includegraphics[width=16cm]{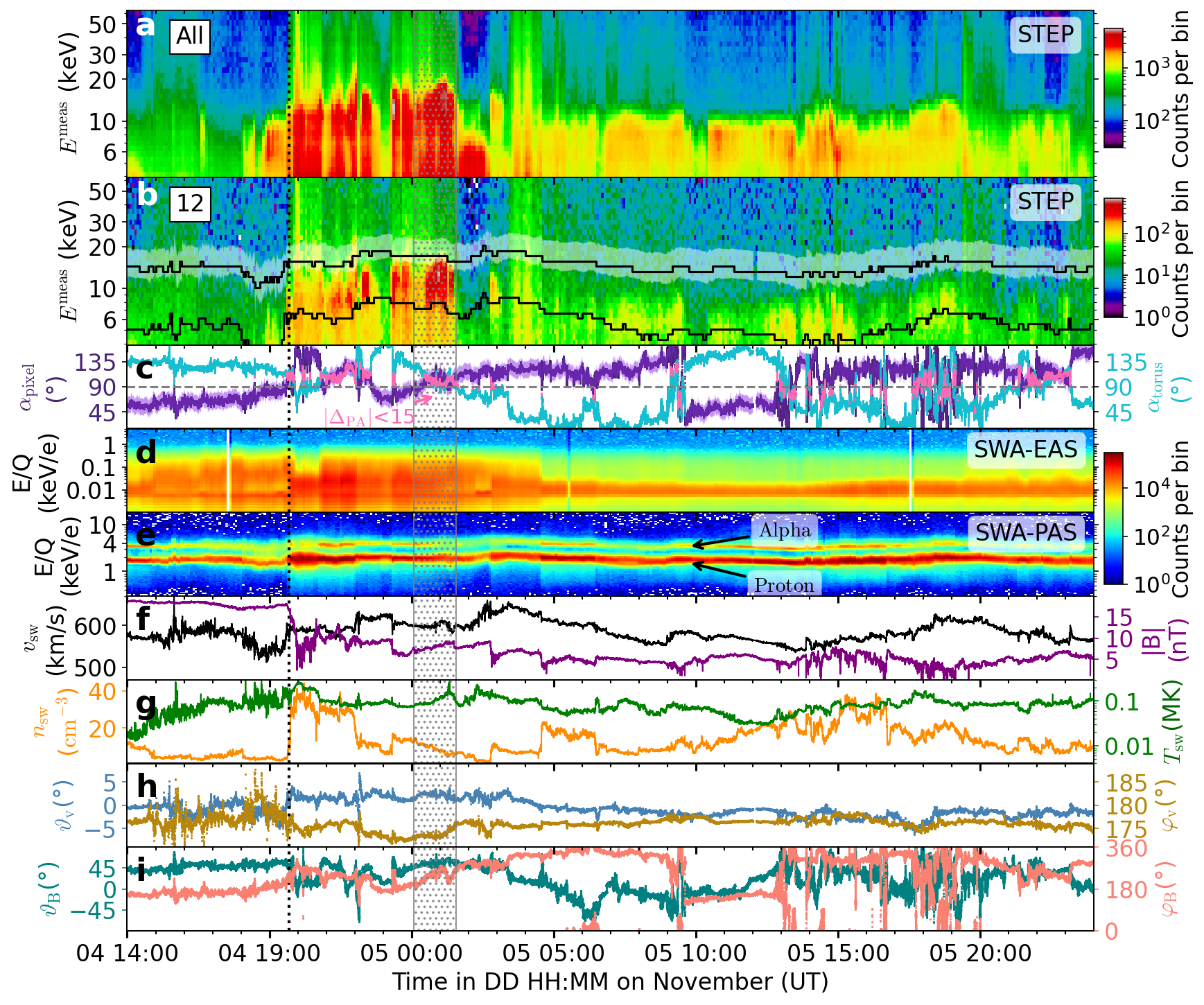}
    \caption{Solar Orbiter observations from 2021 November 04, 14:00 UT to 06, 00:00 UT.
    November 04, 19:42 UT is marked by the dotted line. November 05, 00:04 to 01:34 UT is stippled.
    \textbf{a,} Energy spectrum at 5-minute resolution from STEP/IC, summed from all pixels. 
    \textbf{b,} Same as \textbf{a} but only pixel 12. The white band is the primary energy range of a shell distribution within this pixel's FOV, and the black lines represent the expected measured energy $E^\mathrm{meas}$ range of a narrow shell or torus VDF (Appendix \ref{sec.signatures}).
    \textbf{c,} The PA of the central viewing direction of pixel 12 ($\alpha_{\mathrm{pixel}}$) with $\alpha_{\mathrm{pixel}}\pm15^\circ$ shaded in light purple, and PA of torus  ($\alpha_{\mathrm{torus}}$). Small PA differences ($|\dpa|<15$) are highlighted in magenta (Appendix \ref{sec.signatures}).
    \textbf{d,} Summed energy spectrum at 5-minute resolution from EAS1 and EAS2.
    \textbf{e,} Summed Energy spectrum at 5-minute resolution from PAS's whole FOV. 
    \textbf{f,} The solar wind speed $v_{\mathrm{sw}}$ and magnetic field strength $B$ in the SRF.
    \textbf{g,} Solar wind density $n_{\mathrm{sw}}$ and temperature $T_{\mathrm{sw}}$.
    \textbf{h,} $\bm v_{\mathrm{sw}}$ elevation angle $\vartheta_v$ and azimuthal angle $\varphi_v$ in the SRF.
    \textbf{i,} $\bm B$ elevation angle $\vartheta_B$ and azimuthal angle $\varphi_B$ in the SRF.}
    \label{fig.overview}
\end{figure*}
In this section, we first present an overview of the Solar Orbiter observations of a 34-hour period \change{that} starts on 2021 November 4, 14:00 UT. when Solar Orbiter was at a radial distance $R=0.85~\mathrm{au}$, longitude $\lambda = 41.9^\circ$, and latitude $\beta =-2.0^\circ$ in ecliptic coordinates.
Then we zoom in on a 90-minute time interval to show detailed STEP observations that show pristine PUI torus distributions in Sect.~\ref{sec.90min}.
Three one-minute intervals are further analyzed to \change{determine He$^+$ torus number densities, and to} estimate the He$^+$ PUI torus injection time \change{proxies} \change{and} injection path length \change{proxies} in Sect.~\ref{sec.1min}.

Figure \ref{fig.overview} provides an overview of the time period of 2021 November 4, 14:00 UT to November 6, 00:00 UT. 
It is highly structured and part of a complex ICME event which started earlier with a shock on November 3 at 14:04 UT \citep{2023MNRAS.520..437T}.  
The sudden change in magnetic field strength and proton density on November 4 at 19:42 UT (panels f and g, dashed lines) marks the end of a listed ICME  \citep{2026ApJ..1001...70M}. 
This event is followed by a compression region which likely corresponds to the interface between the first and a following second ICME-like structure \change{(see, e.g. \citet{2006SSRv..123..177W} for typical  ICME properties)} that \change{has not been} previously identified. 
However, the magnetic field showed a slow rotation for several hours after the interface (panel i), which is a typical signature of a magnetic flux rope.

Embedded in this magnetic flux rope, starting right at the interface on November 4 at 19:42 UT to 01:34 UT the next day, the FOV-integrated STEP observations (panel a) display a clear plateau-like enhancement of suprathermal particles (panel a) up to 20 keV.
In the data of STEP pixel 12 (panel b) the enhancement is also visible.
But in contrast to the FOV-integrated data, clear pronounced peaks are visible.
These peaks are located within the two solid black lines that indicate the expected energy range of a PUI torus or shell signature derived from the local magnetic field and solar wind data (\change{Sect.~}\ref{sec.4.1}).
Towards higher energies the peaks have a sharp cut-off and are smeared out towards lower energies.
Panel c shows the PA of the central viewing direction of pixel 12\change{,} $\alpha_{\mathrm{pixel}}$, \change{and the PA of the torus} $\alpha_\mathrm{torus}$ expected under local plasma conditions (\change{Sect.}~\ref{sec.4.2}).
On this pixel, the coincidence of small differences between the two PAs and the observed signatures in panel b fit the expectations for locally injected PUIs that form a torus VDF remarkably well.

In panel d, the SWA-EAS \change{data} show an enhancement of electrons up to a few hundred eV. In EAS data from a year centered on November, 2021, such an enhancement is unusual and the count rates in this energy range are in the highest decile for similar solar wind speeds.
In the SWA-PAS energy spectra, solar wind protons and alpha particles show clear tracks (panel e). 
Although PAS measures He$^+$ over a wider FOV that includes STEP's FOV and up to the velocities expected for PUIs, no clear signatures of He$^+$ are observed by PAS.
Considering the \change{long injection time proxy required to produce a single count in PAS $\tage_{\mathrm{PAS}}(1)$} (Sect.~\ref{sec.1count} and Table \ref{tab.tage}) this is not surprising since PAS was designed to measure the solar wind bulk and not minor ions.
Magnetic field fluctuations and signatures of wave activity that are ubiquitous in the solar wind are on typical levels in our time period of interest after the dashed line (not shown here).

To differentiate between a torus and a shell VDF in STEP, angular resolution is needed. Therefore, in the following, we investigate individual data of all 15 pixels of STEP. 
The time variability in both panels a and b, and the differences between the two panels demand a higher temporal resolution than the 5 minutes shown here.  
Therefore, we selected a 90-minute interval (stippled) that  includes a larger substructure within this magnetic flux rope where the enhanced signals in STEP are pronounced for further analysis, and consider all STEP pixels individually.

\subsection{Signatures of torus VDFs}\label{sec.90min}
\begin{figure*}
    \centering
    \includegraphics[width=14cm]{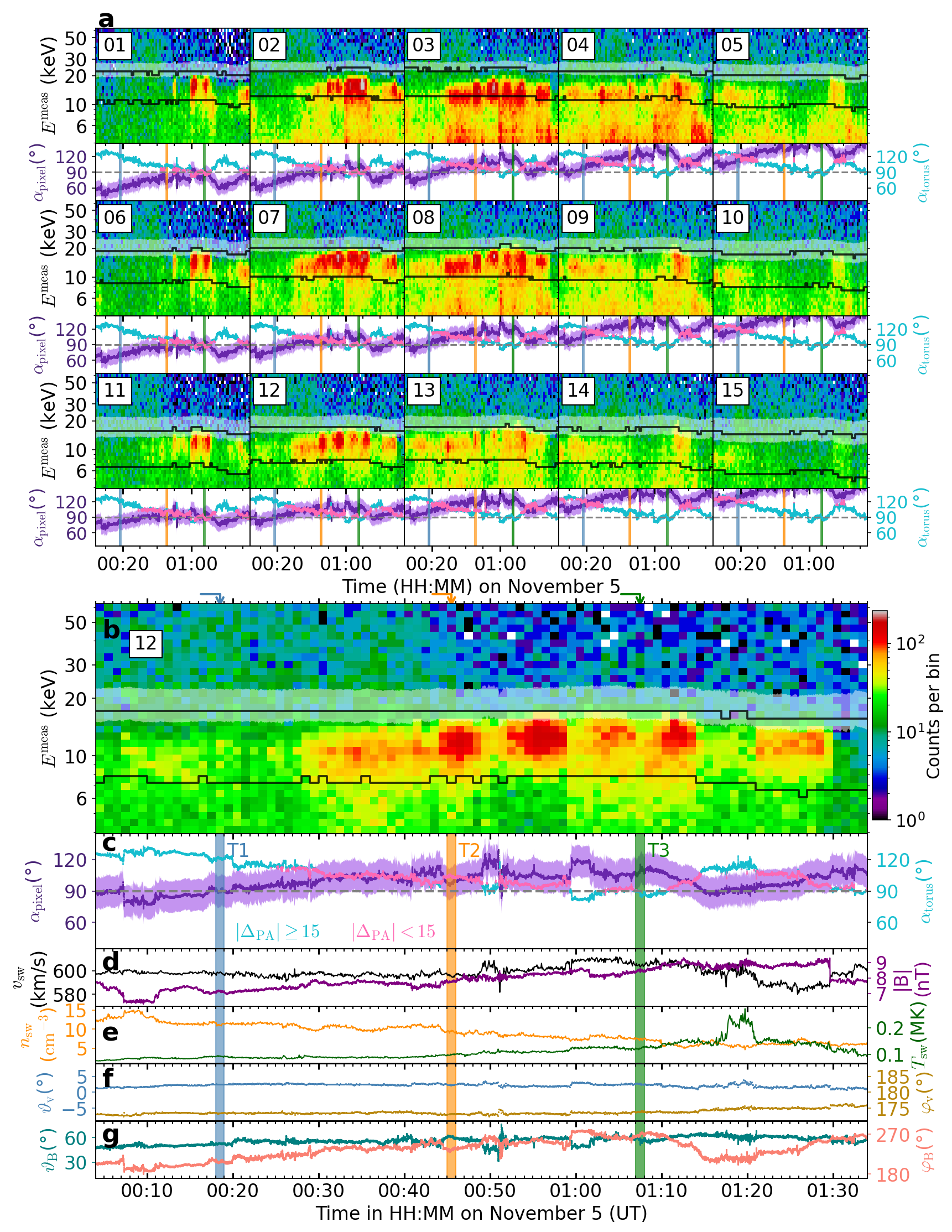}
    \vspace{-0.1cm}
    \caption{Solar Orbiter observations from 2021 November 05, 00:04 to 01:34 UT.
    \textbf{a,} One-minute-integrated energy spectra acquired in the 15 STEP/IC pixels. The white band and black lines represent the primary energy and measured energy $E^\mathrm{meas}$ ranges for $\mathrm{He^+}$ with a shell or torus VDF (\change{Sect.}~\ref{sec.signatures}).
    The central viewing direction PA ($\alpha_{\mathrm{pixel}}$) and $\alpha_{\mathrm{pixel}}\pm15^\circ$, and torus PA ($\alpha_{\mathrm{torus}}$) are given in the panels below.
    \textbf{b-c,} A zoom-in of pixel 12 from \textbf{a}. Three color-shaded minutes start at 00:18 (T1), 00:45 (T2), and 01:07 UT (T3) are selected for further investigations.
    \textbf{d-e,} Solar wind conditions in the SRF. 
    \textbf{f-g,} Elevation angles $\mathrm{\vartheta_v}$, $\mathrm{\vartheta_B}$ and azimuthal angles $\mathrm{\varphi_v}$, $\mathrm{\varphi_B}$ of $\bm v_{\mathrm{sw}}$ and $\bm B$ in the SRF.}
    \label{fig.90min}
\end{figure*}

Figure \ref{fig.90min} shows a zoom-in of the stippled time period from Fig.~\ref{fig.overview}.
Figure \ref{fig.90min}d-g reproduce the same parameters as Fig.~\ref{fig.overview}f-i.
During this period, the solar wind speed (panel d) was high and stable within 580 km/s to 612 km/s, the solar wind density (panel e) was decreasing gradually and over most of the 90 minutes the magnetic field strength overall gradually increased until a sharp discontinuity at the end (panel d).
The magnetic field direction (panel g) was changing smoothly, with a few rapid variations, but generally pointing upwards in the SRF.

Figure \ref{fig.90min}a shows the dynamic energy spectra of this time period at one-minute resolution recorded by each of the 15 pixels, along with the pixel and torus PAs ($\alpha_\mathrm{pixel}$ and $\alpha_\mathrm{torus}$, see \change{Sect.}~\ref{sec.4.2}). The PA of locally produced PUI torus VDFs is inside the FOV of at least one of STEP's pixels in the whole period. The peaks observed on pixel 12 in 5 minute resolution in Fig.~\ref{fig.overview}, appear in Fig.~\ref{fig.90min} on several pixels but not on all pixels in each minute. Overall the peak signatures coincide with small $| \dpa|<15^\circ$. 
The time resolution of 1 minute reveals that these signatures \change{are pronounced with more than 100 counts per energy channel and} change on this time scale.
On the individual pixels the peaks are observed within the expected energy range for a PUI torus or shell VDF.
The shape of the signatures is similar to the example of pixel 12 discussed in the context of Fig.~\ref{fig.overview}, i.e. a sharp cut-off towards higher energies and a smearing out towards lower energies is observed.
As described in Appendix~\ref{sec.aging}, pixels 2, 3, and 4 are visibly affected by instrumental aging, \change{which reduces the measured energy compared to the primary energy, as seen in the energy distributions that are shifted towards lower energies relative to the other pixels.}

In summary, the observed signatures on all 15 pixels have the following characteristic: they do not appear on all pixels at the same time but coincide with small PA differences  $\Delta_\mathrm{PA}$ on the individual pixels, and they display peaks within the energy range of locally produced He$^{+}$ PUI torus or shell VDFs. Thus, we can already conclude that the observed signatures can be attributed to He$^{+}$ PUIs, forming torus VDFs that have not been scattered into a shell distribution yet.
A detailed discussion follows in Sect.~\ref{sec.discussion}.
\change{W}e quantify torus properties for three selected minutes in further detail in the next subsection. 
\subsection{Properties of observed PUI torus VDFs}\label{sec.1min}
\begin{figure*}
    \centering
    \includegraphics[width=13.5cm]{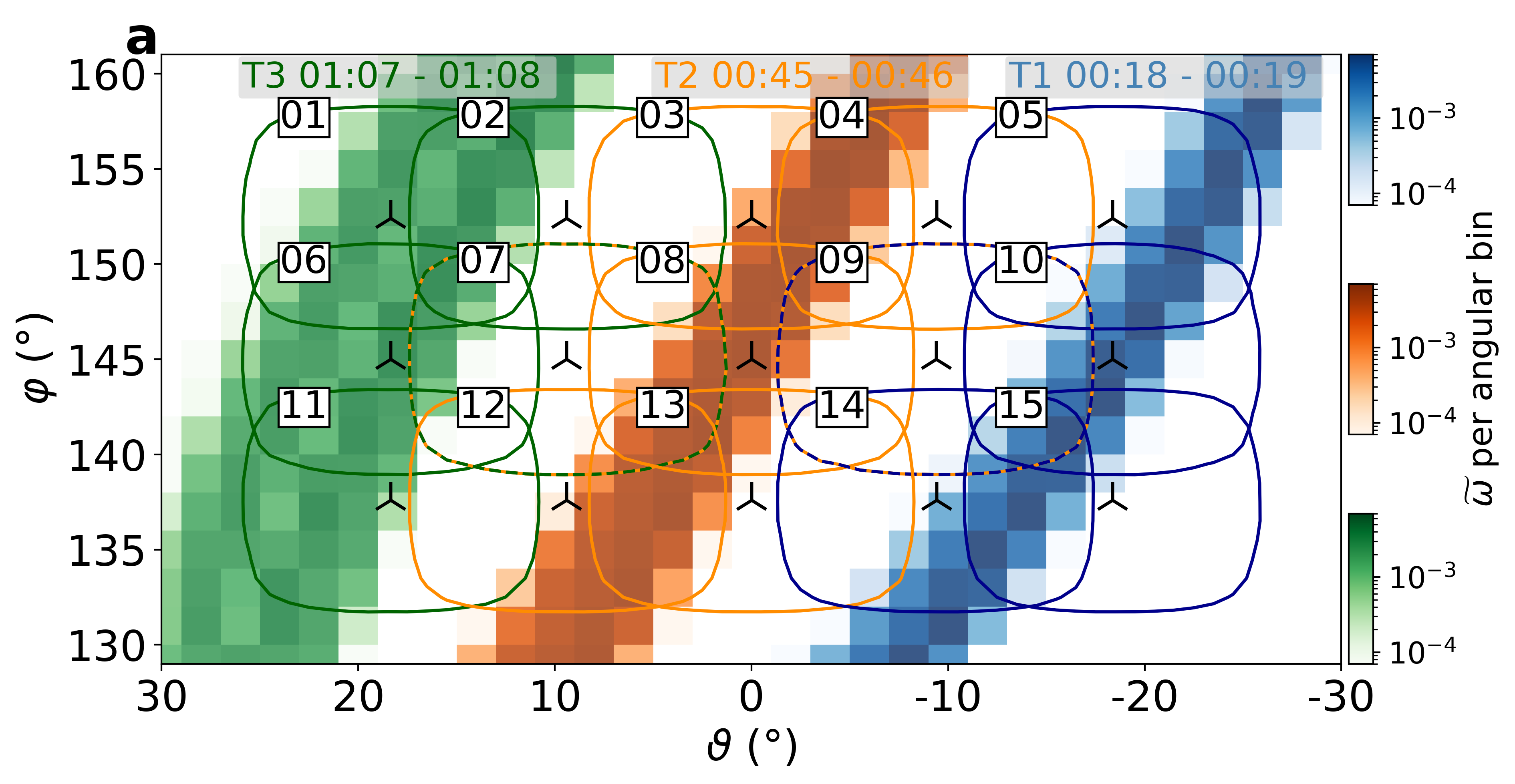}\\
    \vspace{-0.05cm}
    \includegraphics[width=13.5cm]{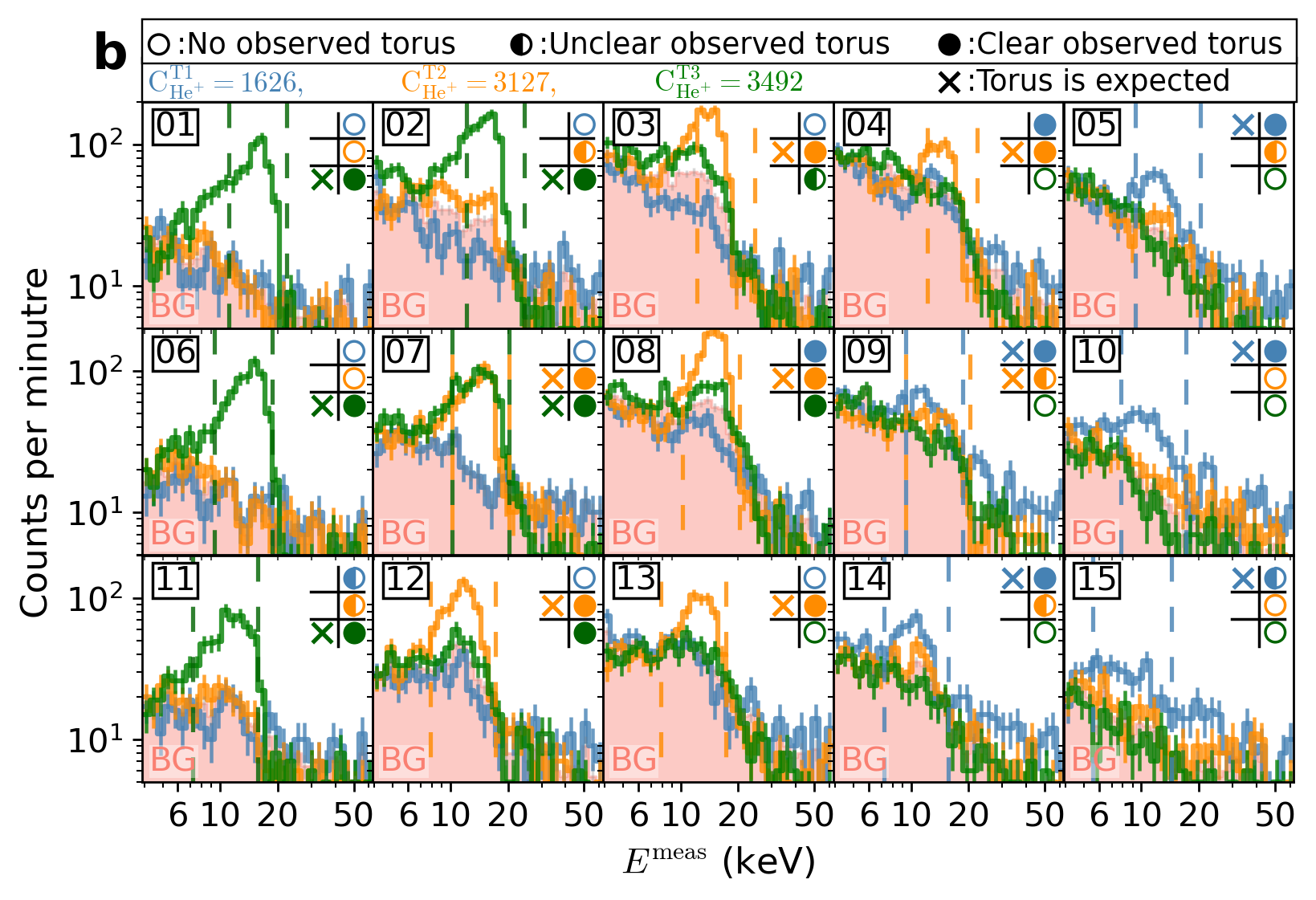}\\
    \vspace{-0.0cm}
    \caption{Torus positions and energy spectra for three selected minutes. 
    \textbf{a,} The expected torus positions based on variations in the 8 Hz magnetic field data are plotted over STEP's FOV in the SRF with elevation angle $\vartheta$ and azimuthal angle $\varphi$. For each minute and each bin the contained averaged fraction $\widetilde{\omega}$ of the expected torus is shown.
    Boundaries represent 5$\%$ of the pixel’s maximum angular response, the color(s) assigned to these boundaries denote  the specific torus (or tori) expected on the respective pixel (\change{Sect.}~\ref{sec.signatures}).
    \wye symbols indicate central viewing directions. 
    For each minute (marked on top), $\widetilde{\omega}$ within each $2^{\circ}\times2^{\circ}$ angular bin is color coded. 
    \textbf{b,} The energy spectra for the three selected minutes (color-matched to \textbf{a}) together with the estimated background spectra (BG, in pink).
    The dashed lines mark the respective $E^\mathrm{meas}$ range, expected for $\mathrm{He^+}$ with a shell or torus VDF.
    Background-subtracted PUI counts for T1, T2, and T3 are given at the top of panel \textbf{b}. 
    Crosses appear when the torus signature is expected by a pixel (\change{Sect.}~\ref{sec.signatures}),
    three fill levels of the circles indicate the subjective clarity of the observed torus signature.
    }
\label{fig.1minute}
\end{figure*}
To further investigate the relation of expected torus position and torus signatures observed on STEP's pixels shown in Fig.~\ref{fig.90min} and to quantify the torus signatures we selected three minutes, T1 (00:18--00:19 UT), T2 (00:45--00:46 UT), and T3 (01:07--01:08 UT), respectively (blue-, orange-, and green-shaded).
\change{There, mean $v_{\mathrm{sw}}$ are found to be [-593.56, 67.29, 26.20], [-591.04, 70.30, 24.62], and [-600.67, 63.89, 27.76] km/s.}
As described in \change{Sect.}~\ref{sec.signatures}, 
\change{the position in velocity space of} a torus that forms under local conditions can be calculated based on the in situ $\bm v_{\mathrm{sw}}$, $\bm B$, and  $\bm v_{\mathrm{inj}}$.

In Fig.~\ref{fig.1minute}a expected torus positions in STEP's FOV are shown together with the angular coverage of the individual pixels. Therein, \change{the $\tilde{\omega}$ as described in Sect.~\ref{sec.4.2} is calculated} based on $\bm B$ in 8 Hz resolution \change{and $\bm v_{\mathrm{sw}}$ in 4 s resolution.
Each colored band thus is the combination of all possible torus positions from that minute, and the width of the band indicates the variability of the solar wind conditions.} 
In each of the three minutes these tori are expected to be well within STEP's FOV, but to be observed on different pixels.
Due to the upwards pointing $\bm B$ the overall situation is similar to \change{one} example (Torus 1) discussed in \change{Sect.}~\ref{sec.4.1}.
In turn, the shape of the observed energy spectra, shown in Fig.~\ref{fig.1minute}b in the respective colors, is also similar to that example.
In the observations the PUI torus signatures are superposed on signals of other suprathermal particles (Appendix~\ref{sec.totalcts}), but clear peaks with a sharp cutoff as expected for torus VDFs can be seen. 
These peaks (marked by filled circles) appear predominantly on the pixels where they are expected (marked by an $\times$ symbol). 
All pixels where clear and most pixels where unclear peaks (half-filled circles) are observed but not expected are adjacent to the pixels that expect tori.

For T1 and T3, the expected torus positions and respective observed torus signatures are located at the edges of STEP's FOV, thus no estimate of the angular width of the torus can be made. For T2, the situation is different and the angular width of the torus is completely contained within STEP's FOV (Appendix~\ref{sec.toruswidth}). We give an upper estimate for the full width of the observed torus of $\gamma\leq20^\circ$ (see also Table \ref{tab.dens}).
Thus, T2 unambiguously represents a narrow torus signature that is restricted to a small angular range.
The numbers of counts that can be attributed to the observed He$^+$ PUI torus signatures, have been estimated by removing the respective backgrounds (pink) of other suprathermal particles as described in Appendix~\ref{sec.totalcts}. From this we also derive \change{He$^+$ number densities}, and estimated injection time \change{proxies}, injection path length \change{proxies} which are summarized in Table \ref{tab.dens}.
The related issues and interpretations are discussed further in Sect.~\ref{sec.discussion}.

\begin{table*}[th]
\caption{\change{Observed PUI number of counts, and proxy properties of three selected tori. }}
\label{tab.dens}
\begin{tabularx}{\textwidth}{lXXXXXXXX}
\hline\hline 
 & \multicolumn{2}{c}{Direct observations}
 & \multicolumn{3}{c}{Virtual detector/ Full responses}
 & \multicolumn{3}{c}{Eqs.~\ref{eq.ntorus2}, \ref{eq.tage}, and \ref{eq.linj}/  Averaged responses}\\
\midrule
VDF & $\ctstorus$ & $\gamma$ & $\niontorus$ & $\tage$ & $\linj$ & 
$\niontorus$&
$\tage$ &  $\linj$\\
 & & ($^\circ$) & ($\mathrm{m}^{-3}$) & (h) & (au) &  ($\mathrm{m}^{-3}$) & (h) & (au)  \\
\hline 
T1  & $1626$ & -- & $1.15 \cdot 10^2$& 15 & $0.22$ & $0.91 \cdot 10^2$&12 & $0.17$  \\
T2  & $3127$ & $\leq 20$ & $1.72 \cdot 10^2$ & 23 & 0.33 & $1.69 \cdot 10^2$& 22 & $0.32$  \\
T3  & $3492$ & -- &$2.64 \cdot 10^2$&35&0.51& $2.33 \cdot 10^2$ & 31 & $0.45$  \\
\hline 
\end{tabularx}
\end{table*}

\section{Implications of observed PUI properties} 
\label{sec.discussion}

PUIs are continuously injected into the outward expanding solar wind plasma. 
As a result the in situ observed VDFs of PUIs are a complex product of the whole history of the in situ solar wind plasma, i.e. they can not be interpreted only in the context of in situ observed plasma conditions.
For the observations we report on in this study the situation appears different, throughout a prolonged period of $\approx 6$ hours we find torus VDFs that resemble the expectations for PUIs injected under the in situ plasma conditions. 

Based on \citet{2013JGRA..118.1389G}, we conservatively expect total He$^{+}$ number densities to be in the range $10^{2}$ m$^{-3}$ -- $10^{3}$ m$^{-3}$ for our period and location of interest.
Compared to these reference values even our lower estimates of torus densities (see Table \ref{tab.dens}) are remarkably high. This suggests, that we observe a significant fraction of the total He$^{+}$ number density in a stable torus,  which supports the inference of very weak PA scattering and \change{could be interpreted as} prolonged injection periods under very similar solar wind conditions. 
\change{Table \ref{tab.dens} provides two estimates for each quantity, both are derived from the observed counts, but the first set  which is used in the following,  considers the full instrumental responses whereas the second set is based on FOV-averaged responses.}

Under the assumption of constant local PUI production rates, prolonged injection periods $\tage$ longer than ten hours (see Table \ref{tab.dens}) would be necessary to explain the observed number densities.
\change{Correspondingly, long injection paths $\linj$} suggest that the observed PUIs \change{would} have been injected over a radial distances between $0.22~\mathrm{au}$ and $0.51~\mathrm{au}$. 
On these spatial scales the assumption of constant neutral He number density $\nn$ \change{should} clearly be violated. 
In addition, the assumption that in equation~\ref{eq.intexp} the quadratic scaling of the total ionization rate cancels with the quadratic scaling by expansion is also likely violated because the electron impact ionization is expected to scale differently \citep{1989A&A...224..290R,2004A&A...426..845G}. 

\change{Further,}
compared to EAS observations of a whole year the unusually high electron count rates shown in Fig.~\ref{fig.overview}d indicate that the contribution of electron impact ionization to the total ionization rates were likely higher than assumed in Sect.~\ref{sec.cts}.
However, even with a generous estimate of ten times higher electron impact ionization rates, i.e about twice as high total ionization rates \citep{2015ApJS..220...27S}, the implied injection periods are still astonishingly long.
\change{For more realistic assumptions a localized model of the He$^+$ PUI densities based on respective local neutral helium densities and local total ionization rates for all locations in the heliosphere that contribute to the trajectory of the in situ observed solar wind parcel would be required which is out of the scope of this study.}

This raises the question how such a narrow \change{densely populated,} and stable torus can develop and be maintained over \change{extended, non-local} scales.
The identification of narrow torus VDFs directly evidences the absence of well populated shell VDFs, i.e. the PA scattering on these scales must have been very weak. 
To what degree an older fully scattered and already cooled PUI sphere population contributed to the non torus background or exists outside STEP's FOV can not be concluded from the observations. 

The observed narrow torus VDFs also imply that $\bm{v}_{\mathrm{sw}}$ and $\bm{B}$ have been very stable on \change{extended} temporal and spatial scales.
Even if the direction of $\bm{B}$ was stable over longer injection periods, i.e. over extended radial distances, the expansion of the plasma would lead to a decrease of $B$ and thus to PA focusing towards field (anti-)parallel PAs. 
For example, the resulting PA focusing over an injection path length $\linj=0.3$, i.e. close to the value estimated for T2 in Table \ref{tab.dens}, would already have broadened the torus width by up to $36.9^\circ$ (Appendix~\ref{sec.pafocussing}). However, this is not observed in the case of T2 where the upper limit of the full width of the observed torus is only half of this expected PA broadening. 
This could \change{hint at} a counteracting process. 

As mentioned in \change{Sect.}~\ref{sec.T_inj.L_inj}, for PA $\neq 90^\circ$ the guiding center lags behind the solar wind bulk, in the case of T2 with a difference velocity  $\bm{v}_\mathrm{diff}=[ 40, 59, -128 ]$~km/s. Thus, the guiding center of these particles drift backwards, i.e. antiparallel to the magnetic field, through the observed 3D magnetic structure.
In our case, this potentially leads to an accumulation of PUIs towards the end of the substructure, i.e. at about 01:30 UT in Fig.~\ref{fig.90min}. As a result, the particles experienced a combination of the negative gradient of the magnetic field due to the expansion of the structure and the positive gradient within the structure.  We hypothesize that these focusing processes towards and away from 90$^\circ$ here might have counteracted each other and stabilized the torus signature at T2.

Finally, we put the STEP observations in the context of expectations for other active PUI instruments as exemplified by the PLASTIC \citep{Galvin2008} instrument of STEREO-A. Unlike STEP, PLASTIC identifies ion species. However for PLASTIC this comes with the price of smaller expected PUI count rates. 
\change{Under the assumption that a He$^+$ PUI VDF is only populated by the locally injected PUIs, we derived in Sect.~\ref{sec.cts} the expected injection time $\tage_\mathrm{PLASTIC}(1)$ (see Table \ref{tab.tage}) required  to observe a single He$^+$ count with STEP and PLASTIC, respectively. For PLASTIC, this expected injection time is $\sim 3.5$ hours. Nevertheless, previous studies, e.g. \citep{2015A&A...575A..97D, 2025A&A...696A.115K, 2025ApJ...981...35O} have observed a clearly identifiable torus feature inside and outside of the He$^+$ focusing cone. Although this was not previously discussed, this implies that also PLASTIC has observed more He$^+$ PUIs in torus VDFs than expected from local injection alone. Thus, these previous studies are also consistent with He$^+$ PUI torus VDFs that remain stable for sufficiently long time periods to accumulate PUIs over extended temporal and spatial scales and correspondingly weak scattering.}  
\change{For STEP the expected injection time $\tage_{\mathrm{STEP}}(1)$ is with $\sim 33$s much shorter than for PLASTIC. Therefore, it is  not surprising that STEP observes considerably more He$^+$ counts than PLASTIC. The high count rates allow to investigate the properties of He$^+$ PUIs  at much shorter time scales which in turn makes the stability of densely populated He$^+$ PUI torus VDFs easier to detect.}

\section{Conclusions}\label{sec.conclusion}
During an extended period of about 6 hours starting at 19:42 UT on November 4, 2021, STEP observed enhanced count rates in highly anisotropic signatures.
We find that throughout this period these signatures can be clearly attributed to interstellar He$^{+}$ PUI torus VDFs that appear to have been injected under \change{the} in situ observed plasma conditions.
These torus VDFs are variable at a one minute time scale. 
We can compare our observed torus densities to expectations from local fresh injection on the one hand and on the other hand to the total He number density of all PUIs injected into the solar wind up to the spacecraft location. 
We find with the former that the observed torus densities are very high compared to local expectations and with the latter that they are comparable to the total He$^+$ PUI number densities. 
Both independently imply that the observed He$^+$ PUIs require extended periods of PUI production, i.e. the observed He$^+$ PUIs have been injected over extended radial distances \change{that can no longer be considered as local.}

A second important observation is that the observed PA widths of the tori are narrow and thus show no evidence of significant PA scattering. 
In addition, the injection over such long radial distances without PA scattering should have resulted in significant broadening of the torus due to magnetic focusing from a negative $B$ gradient in the expanding solar wind plasma. 
However, we find no  such broadening. 
This in turn might point towards a more local PUI production, which would require orders of magnitude higher ionization rates than typically reported in the literature \citep{2020ApJ...897..179S,2022ApJS..259...42S, 2025ApJ...995...57A}.
In fact EAS observations show enhancements of suprathermal electron densities, that would result in higher local electron impact ionization rates, but the effect on the total ionization rates would be well below one order of magnitude even if the local electron impact ionization was 10 times higher. Therefore, we suggest that higher ionization rates are not sufficient to explain our observations. 

In the absence of strong PA scattering, for PA $\neq 90^\circ$ the guiding center of the PUIs does not comove exactly with the solar wind bulk. 
Instead PUIs drift backwards through substructures which are convected with the solar wind bulk. 
In our case an observed positive gradient in the local substructure and the negative gradient due to expansion could have canceled each other out.
\change{A potentially long-lasting} relative motion of the guiding center with respect to the solar wind bulk, \change{which results from a long scattering mean free path,
is important and should be considered in particle transport models \citep{2010SpWea...8.0E02S,2015JGRA..120.9269C,2026SSRv..222....6C}. 
However, even models that consider this effect rely on additional assumptions on remote solar wind conditions. These assumptions are not directly verifiably by any measurement which inhibits the comparison of model results with observations.}

Our study \change{provides a} first glance \change{at} the complexities of PUI physics that have been unlocked by the unprecedented high time resolution and uninterrupted velocity space coverage of STEP.   
Recently, in September 2025, \change{NASA's Interstellar Mapping and Acceleration Probe ~\cite[IMAP;][]{2025SSRv..221..100M} was launched. The Solar Wind and Pickup Ion~\cite[SWAPI;][]{2025SSRv..221..108R} Instrument and the Compact Dual Ion Composition Experiment~\cite[CoDICE;][]{2025SSRv..222....1L} on board IMAP will provide valuable PUI measurments, including 3D VDFs.
}
Together, Solar Orbiter and IMAP may advance the understanding of processes related to the pickup process and VDF evolution.

\begin{acknowledgements}
Solar Orbiter is a mission of international cooperation between ESA and NASA, operated by ESA. This work was supported by the German Federal Ministry for Economic Affairs and Energy and the German Space
Agency (Deutsches Zentrum für Luft- und Raumfahrt, e.V., (DLR)), grant number 50OT2002.
Chaoran Gu is supported by the German Research Foundation grant number WI 2139/12-1.
Verena Heidrich-Meisner is supported by DLR grant number 50OC2104.

        \end{acknowledgements}

\bibliographystyle{aa} 
\bibliography{reference} 

\appendix
\onecolumn

\section{Virtual STEP detector}\label{sec.vd}

We developed a virtual STEP detector to (1) visualize the torus and shell VDF signatures in \change{Sect.}~\ref{sec.signatures}, (2) estimate the expected number of $\mathrm{He^+}$ PUI counts and $\mathrm{He^+}$ densities in STEP in Sect.~\ref{sec.cts}, and (3) compare observed signatures with expected signatures of PUI torus and shell VDFs in Sect.~\ref{sec.results}.
\begin{figure}[ht!]
    \centering
    \includegraphics[width=14cm]{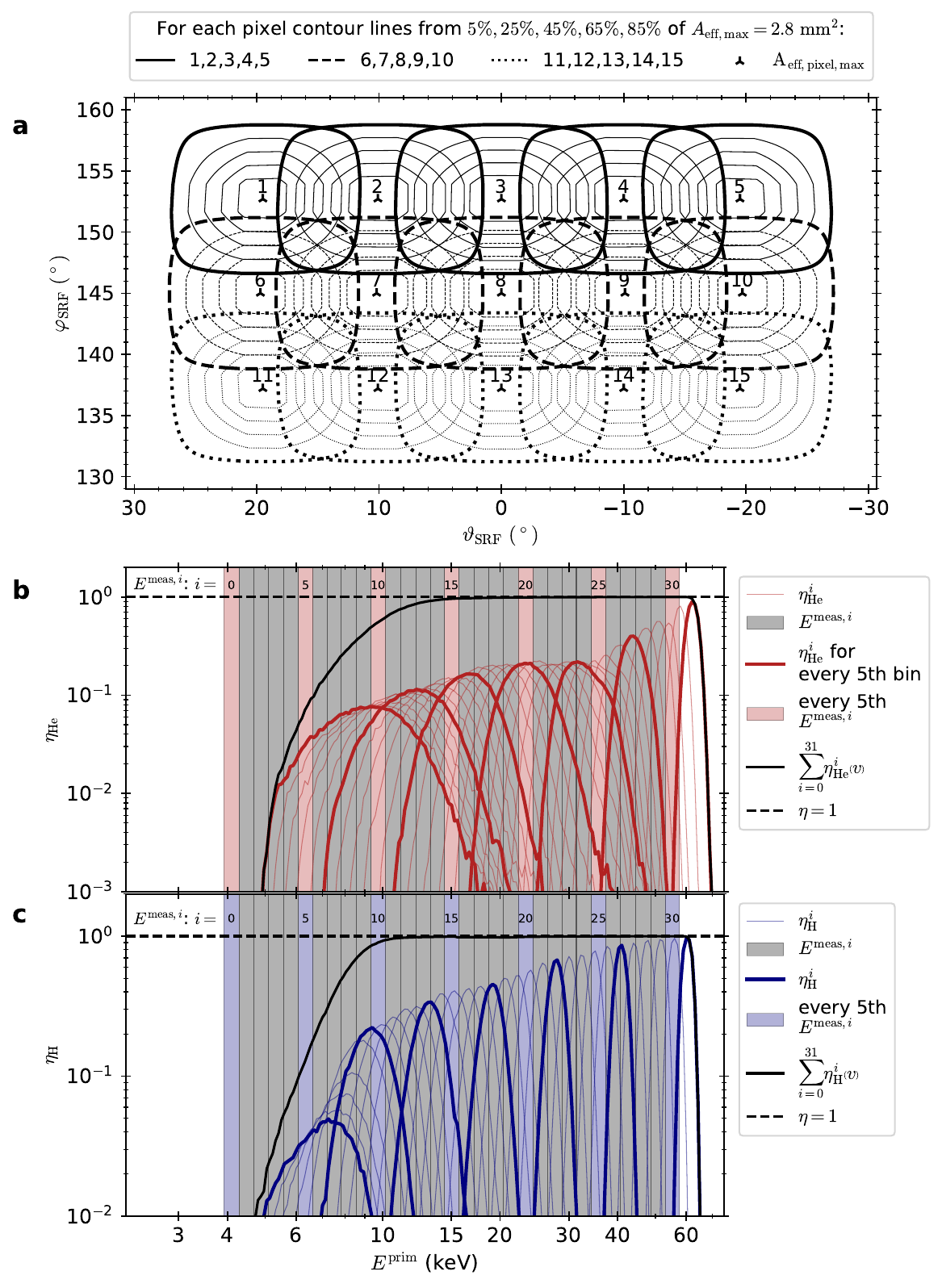}
    \caption{Three-dimensional He \change{and H} response of  STEP/IC. \textbf{a,} The projected effective area $A_{\rm{eff}}$ of the pinhole for each pixel in ecliptic SRF coordinates. The pixel-wise $A_{\rm{eff,pixel}}$ is shown as contour lines for $5\%, 25\%, 45\%, 65\%, 85\%$ of the global maximum $A_{\rm{eff}}=2.8\ \rm{mm}^2$ which occurs in pixel 8. The contour lines for the first five pixels 1-5 are drawn with solid lines, for pixels 6-10 with dashed lines, and for pixels 11-15 with dotted lines. Each pixel is marked with the corresponding number and a $\wye$-symbol for the pixel-wise maxima $A_{\rm{eff,pixel,max}}$. \textbf{b,} The energy calibration of STEP for He obtained from a GEANT4 simulation. \change{ The efficiency $\eta_\mathrm{He}^i$ for each instrumental bin $i$ is shown for each of the $31$ first STEP energy bins. This illustrates in which energy range the individual bin is sensitive in $E^{\mathrm{prim}}$.  Every fifth line is highlighted in red for visibility.} The shaded areas in the background indicate the positions of the 31 first STEP energy bins in measured energy $E^{\mathrm{meas}}$ wherein again every fifth bin is colored red for visibility. \change{The sum of all responses is shown with a solid black line. The dashed vertical black line marks 1. \textbf{c} Same as b, but for protons.}}
    \label{fig.energy.response}
\end{figure}
\change{The virtual STEP detector numerically integrates Eq.~\ref{eq.intinstr} for each pixel and energy bin for an input velocity space density distribution $\delta n(\bm{v})$.}
\change{The instrumental responses for each pixel and energy bin are split} in two parts, angular responses \change{$A(\varphi,\vartheta)$} and energy responses \change{$\eta(E)$} \change{with a duty cycle $\zeta(\bm{v}) = 1$}.
The angular responses for the IC can be obtained by a simple geometric consideration that traces particle trajectories through the pinhole to each pixel \change{and are shown in Figure ~\ref{fig.energy.response}a}. 
The energy responses of STEP's solid-state detector for $\mathrm{He}$ \change{and $\mathrm{H}$} have been modeled using the GEANT4 \citep{2003NIMPA.506..250A} toolkit with a similar setup as for the proton responses published in the Solar Orbiter Archive \change{and are shown in Figure ~\ref{fig.energy.response}b, and c respectively.}
It is clearly visible that the responses are not rectangular boxes and overlap considerably. 
As a result, even a monoenergetic beam will be measured in several energy bins.

\section{Instrumental aging of STEP pixels}\label{sec.aging}

\change{Silicon semiconductor solid state detectors are basically reversed-biased planar diodes.
Traversing particles lose kinetic energy and produce electron-hole pairs and the resulting charge pulse is measured.
For a healthy silicon SSD on average one electron-hole pair is produced per 3.6 eV energy loss and the resulting charge pulse is proportional to the total kinetic energy loss of each particle.
Beside the desired electron-hole pair production, ions can also damage the lattice of the SSD by displacing single atoms or by producing clusters of displacements.
Generally a distinction is made between surface and bulk damage. 
Surface damage refers to damage in the non-active volume of the detector, e.g. the dead layers in front of the active volume.
Bulk damage refers to damage in the active volume of the detector.
In the following, we focus on bulk damage.}

\change{Lattice defects in the active volume introduce transition levels between the valence and the conduction band of the SSD which can effectively trap electrons.
As a result these electrons do not contribute to the generated charge pulse and the measured signal for the same kinetic energy loss is decreased.
These effects are well known and reasonably well understood for high energy applications in well characterized radiation fields, e.g. at the Large Hadron Collider at CERN.
For radiation fields of comparably low energies, that are highly variable and poorly characterized, the damage is poorly understood and very hard to quantify.
In high energy applications the energy of particles is high enough to easily penetrate an SSD producing only very few displacements along their trajectory.
Therefore the damage is spread over the whole detector volume and scales with the total fluence of particles through the detector.
At the lower energy end, particles do no longer penetrate the detector but are stopped and get stuck.
Both the average penetration depth as well as the efficiency to produce a displacement at a certain depth depend on the ion species and primary kinetic energy.
In general the efficiency to generate displacements increases with decreasing ion velocity.
Ions in the energy range of STEP are stopped and get stuck in the very front layers of the active volume ($< 1~\mu$m compared to the $300~\mu$m thickness of the SSD).
In addition they generate a large number of displacements.}

\change{As illustrated in Figure~\ref{fig.1}b in the solar wind frame of reference at the lower energies the FOV of STEP's pixels 3, 8 and 13 are covering different parts of the suprathermal range, with pixel 3 measuring closest to the solar wind bulk.
The overall trigger rates of the 15 pixels, i.e. the fluence the individual pixels have observed, reflect their distance to the average solar wind bulk flow.
Thus, the expected damage or change in the calibration is time dependent, different on each pixel, and is strongest on pixel 3.}

\change{This instrumental aging affects all STEP observations. 
In dispersive ion Solar Energetic Particle (SEP) events it can be seen that ions appear to arrive early with a systematic pattern over all 15 pixels.
This effect is increasing with mission time, i.e. with increasing fluence.
Dispersive electron SEPs do not show signs of changing calibration of the individual pixels and electrons have much higher penetration depths. 
Thus, the conclusion that the damage is restricted to the front layers of the active detector volume is supported.
Efforts to quantify the fluence-dependent aging with the aim to provide time dependent calibration for the individual pixels are ongoing.
The observations we report on are made at an early stage of the mission, when a visible change in the energy response is restricted to pixels 2, 3 and 4.
}

\section{\change{He$^+$ and H$^+$ torus  and shell VDFs in STEP}} \label{sec.identification}
\begin{figure}[ht!]
\centering\includegraphics[width=14cm]{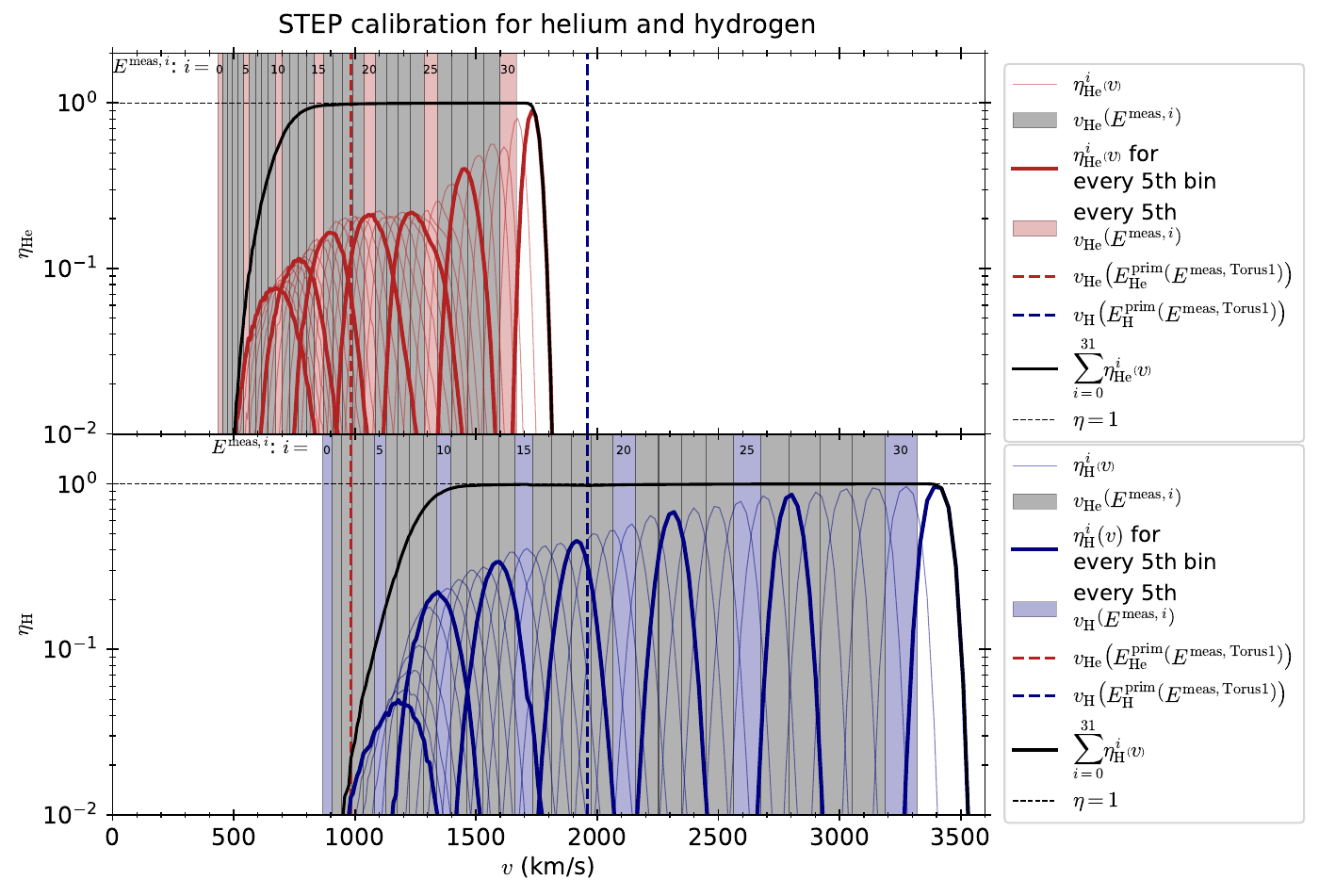}
\caption{\label{fig.velocity_responses}\change{One-dimensional He and H velocity responses for STEP in a similar format as the corresponding energy responses in Fig.~\ref{fig.energy.response}b and c.}}
\end{figure}

\begin{figure}[ht!]
    \centering
    \includegraphics[width=0.95\linewidth]{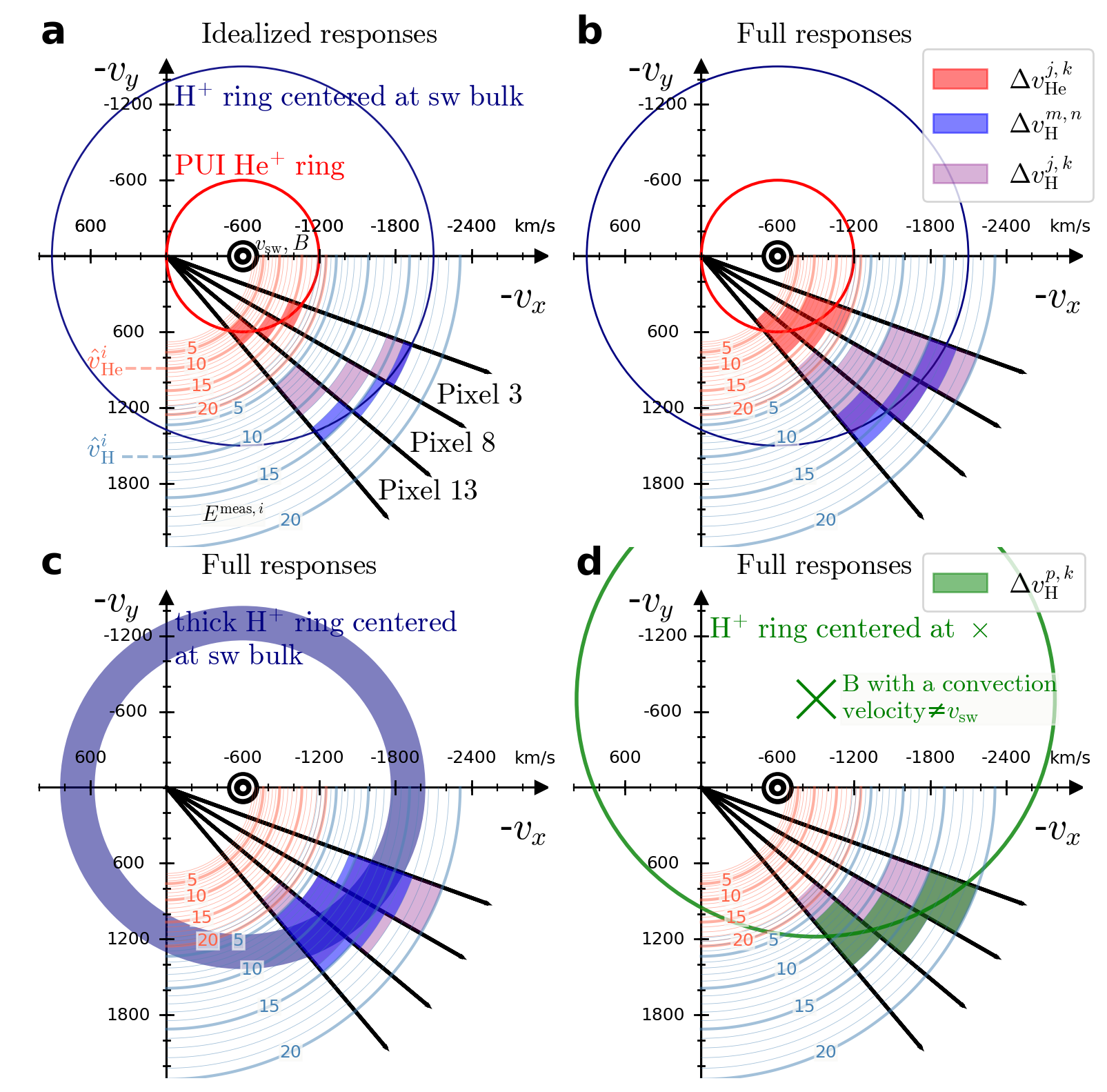}
    \caption{\change{Sketches of different ring VDFs in velocity space in the SRF, and the corresponding instrumental responses (idealized for \textbf{a}, and full for others) within the FOV of three pixels of STEP.
    As for Torus  1 in Fig.~\ref{fig.1}, the solar wind bulk velocity is $v_{\mathrm{sw}}=-600$ km/s, and the magnetic field is perpendicular to the $v_x$--$v_y$ plane. 
    The average velocities for selected velocity bins (compare Fig.~\ref{fig.velocity_responses}), $\hat{v}^i_{\mathrm{He}}$, in red and, $\hat{v}^i_{\mathrm{H}}$, in blue are shown as 90$^\circ$ arcs, labeled every 5th bin.
    Within one pixel's FOV, a color shaded region, $\Delta v_{\mathrm{X}}^{y,z}$, is the velocity range translated from the instrumental energy responses from bin $y$ to bin $z$ for ion species $\mathrm{X}$ in the respective color.
    Purple illustrates the helium velocity ranges if interpreted as proton velocities. 
    \textbf{a,} Sketch of a He$^+$ ring (the same as Torus 1 in Fig. \ref{fig.1}b) and a H$^+$ ring, both centered at the solar wind bulk, with a radius at $600$ and $1500$ km/s respectively. 
    \textbf{b,} Same as a, but with full responses.
    \textbf{c,} Sketch of a thick H$^+$ ring, centered at the solar wind bulk.
    Purple regions are the same as in b.
    \textbf{d,} Sketch of a H$^+$ ring, but centered at [-900,-700] km/s with a radius of 1880 km/s. Purple regions are the same as in b.
    }}
    \label{fig.sketch.H.ring}
\end{figure}
\change{
STEP was not designed to identify ion species. However, as discussed in Sect.~\ref{sec.signatures} for the special case of a torus or shell VDF, characteristic signatures are observed and expected in STEP for such a He$^+$ PUI VDF. Here, we discuss the differences of signatures of a torus or shell VDF of He$^+$ and H$^+$ in STEP, respectively. 
As shown in Fig.~\ref{fig.sketch.H.ring}a, regarding the energy response, a PUI $\mathrm{He}^{+}$ ring (red) should trigger signals on certain energy bins of pixel 3, 8 and 13 (red shaded).
For protons to trigger a signal in the same energy bins, the corresponding proton velocity (blue shaded) is higher and the pattern of triggered  energy bins is different. As discussed in Sect.~\ref{sec.signatures}, the same narrow torus VDF is observed in different energy bins and over different energy bin ranges for different STEP pixels. For a proton torus to fully mimic the same response as for a He$^+$ torus, this geometric broadening also would need to be reproduced. The velocity range protons would need to cover is illustrated with purple shaded bins. The radius of the H$^+$ torus is chosen to match the upper bound of the purple region on pixel 3. The different underlying velocities for He$^+$ and H$^+$ alone already lead to systematically different patterns of a torus signature on the different STEP pixels (compare blue and purple shaded regions in Fig.~\ref{fig.sketch.H.ring}a). }

\change{However, in this sketch we assumed instrumental responses that are idealized in the sense that they are box-shaped and in particular non-overlapping. As shown in Fig.~\ref{fig.energy.response}b and c for the energy responses and in Fig.~\ref{fig.velocity_responses} as velocity responses, this is not the case for STEP. These responses are taken into account in Fig.~\ref{fig.sketch.H.ring}b-d. In Fig.~\ref{fig.sketch.H.ring}b, realistic instrumental responses increase the width of the signatures of He$^+$ and H$^+$ torus or shell VDFs, that is, the blue, red, and purple shaded regions are wider than in Fig.~\ref{fig.sketch.H.ring}a. In particular, the (blue) signature of a narrow H$^+$ torus is much narrower than the corresponding signature for a He$^+$ torus (blue and purple signatures). Therefore,  in Fig.~\ref{fig.sketch.H.ring}c the width of the proton torus example (dark blue shaded ring) is increased. However a fixed torus energy width is still not sufficient to match the purple signature, i.e., on pixel 3 the purple signature extends to higher energies than the blue-shaded H$^+$ signature, while on pixel 13, the purple signature extends to lower energies than the blue-shaded H$^+$ signature.
This illustrates that while the characteristic patterns of a He$^+$ PUI torus feature and a H$^+$ torus feature can match on single pixels, the full pattern over all pixels is systematically different. Thus, taking the full cutoff pattern into account allows a species determination with STEP  for the special case of torus or shell like VDFs.}

\change{The only option for a H$^+$ torus VDF to mimic the response of He$^+$ torus VDF is illustrated with the green VDF in Fig. \ref{fig.sketch.H.ring}d. However, this hypothetical torus VDF is not centered at the solar wind bulk velocity. Such a situation is unrealistic and unphysical.}

\section{PUI torus counts observed by STEP} 
\label{sec.totalcts}
From Sect.~\ref{sec.cts} we expect a PUI torus signature to be visible only on a subset of STEP's pixels depending on the current solar wind conditions.  As described in Sect.~\ref{sec.instrument}, STEP can not discriminate different ion species.
Other contributions to the observed spectra could be suprathermal tails of solar wind protons and alpha particles, solar energetic event particles, as well as PUIs with shell or sphere VDFs.
Without strong assumptions about the true nature of these contributions, we take a heuristic approach to remove them. 

We observe that the background in the $1$ minute measurements on the individual pixels for the time period of interest shown in Sect.~\ref{sec.1min} is similar whenever no clear torus signature is present.
Hence, in the following, we consider contributions other than the PUI torus VDFs as a constant and isotropic background for this period. So, this background does not change with time and does not depend on the magnetic field direction.
With these assumptions we estimate the observed number of counts $\cts$ in the PUI torus signatures observed in the three selected minutes, T1, T2, and T3, investigated in Sect. \ref{sec.1min}.
The background spectra are estimated individually for each pixel, and are defined as the average of the spectra for which no torus signature is expected under the current solar wind conditions on this pixel. 
For example, in Fig. \ref{fig.1minute}b the background spectrum (pink) for pixel 1 is the mean spectrum of T1 (blue) and T2  (orange).
The background for pixel 7 is represented by the T1 spectrum because the T2 and T3 spectra expect torus signatures. 

This background estimation has some limitations. For example, as visible in Fig.~\ref{fig.1minute}b, in T1 the background is higher on most pixels than in T2 and T3. 
For another example, on pixel 8, a torus signature is visible in the STEP observations for T1, T2, and T3, although this is not expected for T1 and T3. 
In theses cases, our background estimation is at its limit, and for example, in the case of pixel 8 by taking the average of T1 and T3 as the background spectrum, we underestimate the number of counts in the torus signature. 
As a result the background is likely overestimated by our approach. 

With this background estimate, we can simply subtract the resulting background spectra for each pixel and obtain an estimate of the number of PUI counts $\cts$ attributed to the torus from the respective  sum over all pixel where we expect to detect a torus. 
Thus, the resulting observed PUI counts per minute and He$^+$ torus number densities are conservative lower estimates and are included in Table \ref{tab.dens}.
The He$^+$ torus number densities $\niontorus$, the injection path length\change{s proxies} $\linj$, and the injection time\change{s proxies} $\tage$ are discussed in Sect.~\ref{sec.discussion}.
\section{PUI torus width observed by STEP}
\label{sec.toruswidth}
As visible in Fig.~\ref{fig.1minute}, in time period T2 the observed torus signatures are confined within STEP's FOV. 
Pixels 1 and 6 that look at PAs smaller than the expected torus PAs and pixels 10 and 15 that look at PAs larger than the expected torus PAs observe no torus signatures.
Among these pixels, Fig.~\ref{fig.1minute} shows that the lower right edge of pixel 6 and the upper left edge of pixel 10, respectively, look closest to the expected torus positions.
Therefore, we take the sum of the minima of the absolute PA difference of these two pixels and the expected tori PAs within T2 plus the spread of the expected torus PAs as an upper estimate for the full width, $\gamma$, of the observed torus signature (see Table~\ref{tab.dens}).
We emphasize that this is a loose upper limit and any larger width would not be consistent with the STEP observations.

\section{Pitch angle focusing}
\label{sec.pafocussing}
From the conservation of the magnetic moment of charged particles follows a change of the particles' PA if the magnetic field strength $B$ changes.
The resulting PA $\alpha_{2}$ can be calculated from the original PA $\alpha_{1}$ if the magnetic field changed from $B_{1}$ to $B_{2}$ by
\begin{equation}
\label{eq.pafocussing}
\alpha_{2} = \arcsin{\left(\sin{\left(\alpha_{1}\right) \cdot \sqrt{\frac{B_{2}}{B_{1}}}}\right)} \enspace.
\end{equation}
In the outward expanding solar wind plasma $B$ is expected to decrease radially and thus the PA $\alpha_{\mathrm{inj}}$ of PUIs that have been injected closer to the Sun should have changed according to Eq. \ref{eq.pafocussing} towards the parallel or anti parallel direction, for $\alpha_{\mathrm{inj}}<90^{\circ}$ and  $\alpha_{\mathrm{inj}}>90^{\circ}$, respectively.
Close to the Sun, the magnetic field is expected to decrease with $r^{-2}$ \citep{2013LRSP...10....5O}. Thereby we can calculate the expected PA changes of PUIs that have been injected in the history of the in situ plasma (see Eq.~\ref{eq.int}).

For the situation in T2 (see Sect.~\ref{sec.1min}), this focusing effect can be quantified as follows.
For a distance of $\linj =  0.1$~au ($\linj = 0.3$~au), i.e PUIs that have been injected at $r = 0.75~ \mathrm{au}$ ($r = 0.55~ \mathrm{au}$), at $\alpha_{\mathrm{inj}} = 104.2^{\circ}$, i.e. the in situ observed conditions in T2, we obtain an in situ PA $\alpha_{\mathrm{in~situ}}  = 121.2^{\circ}$ ($\alpha_{\mathrm{in~situ}} = 141.1^{\circ}$). 
Thus, the contribution of the PUIs to the joined VDF that have been injected further in should have broadened the torus VDF significantly in PA from $\alpha_{\mathrm{inj}} = 104.2^{\circ}$ towards $180^{\circ}$ for the injection path lengths $\linj=0.33$ (see Table \ref{tab.dens}).

\end{document}